\documentclass[12pt]{article}%
\usepackage{amsmath}
\usepackage{graphicx}%
\usepackage{amsfonts}%
\usepackage{amssymb}
\usepackage{xcolor}
\usepackage{placeins}
\usepackage{enumerate}
\usepackage{natbib}
\usepackage{authblk}
\usepackage{lineno}

\newcommand{\beginsupplement}{%
        \setcounter{table}{0}
        \renewcommand{\thetable}{S\arabic{table}}%
        \setcounter{figure}{0}
        \renewcommand{\thefigure}{S\arabic{figure}}%
     }

\newcommand{\bmu}{ \mbox{\boldmath $\mu$} }

\newcommand{\bzero}{\textbf{0}}

\newcommand{\bs}{\textbf{s}}

\title{Long-term Spatial Modeling for Characteristics of Extreme Heat Events}
\author[1]{Erin M. Schliep}
\affil[1]{Department of Statistics, University of Missouri, Columbia, Missouri, USA\\
{\it schliepe@missouri.edu}}
\author[2]{Alan E. Gelfand}
\affil[2]{Department of Statistical Science, Duke University, Durham, North Carolina, USA.}
\author[3]{Jes\'{u}s Abaurrea}
\author[3]{Jes\'{u}s As\'{i}n}
\author[4]{Mar\'{i}a A. Beamonte}
\author[3]{Ana C. Cebri\'{a}n}
\affil[3]{Departmento de M\'{e}todos Estad\'{i}sticos, Universidad de Zaragoza, Zaragoza, Spain.}
\affil[4]{Facultad de Econom\'{i}a y Empresa, Universidad de Zaragoza, Zaragoza, Spain.}

\begin{document}
\maketitle
\vspace*{0.5cm}

\begin{abstract}

There is increasing evidence that global warming manifests itself in more frequent warm days and that heat waves will become more frequent. Presently, a formal definition of a heat wave is not agreed upon in the literature. To avoid this debate, we consider extreme heat events, which, at a given location, are well-defined as a run of consecutive days above an associated local threshold. Characteristics of EHEs are of primary interest, such as incidence and duration, as well as the magnitude of the average exceedance and maximum exceedance above the threshold during the EHE.

Using approximately 60-year time series of daily maximum temperature data collected at 18 locations in a given region, we propose a spatio-temporal model to study the characteristics of EHEs over time. The model enables prediction of the behavior of EHE characteristics at unobserved locations within the region. Specifically, our approach employs a two-state space-time model for EHEs with local thresholds where one state defines above threshold daily maximum temperatures and the other below threshold temperatures. We show that our model is able to recover the EHE characteristics of interest and outperforms a corresponding autoregressive model that ignores thresholds based on out-of-sample prediction.

\end{abstract}
\bigskip


\section{Introduction}
\label{Sec:Intro}

There is strong evidence of global warming induced as a result of the increasing concentration of greenhouse gases in the atmosphere \citep{Lai19}. It can be expected that this global warming will manifest itself in more frequent warm days and many papers suggest that heat waves will become more frequent \citep{Lemonsu14, Alexander16}. This anticipated scenario along with the exceptionally long heat wave observed in Europe in August 2003 has increased interest in analyzing climate extremes. The analysis of heat waves is particularly important due to the potential for serious anthropogenic, environmental, and economic impacts \citep{Amengual14, Campbell18}. Extreme heat raises significant health concerns in humans as it can result in death, change the range or niche for plants and animals, and lead to heat-driven peaks in electricity demands or lost crop income.

A challenge in analyzing heat waves stems from the on-going debate over its exact definition.
According to the World Meteorological Office (WMO), a period persisting at least three consecutive days of marked unusual hot weather (maximum, minimum and daily average temperature) over a region with thermal conditions above given thresholds based on local climatological conditions can be considered a heat wave.
This definition suggests that analyses of heat waves require temperature series at a daily scale, but offers no operational guidance with regard to the various choices of implementation.
\cite{Khaliq05} and \cite{Reich14} used only maximum temperature, while \cite{Keellings14} and, more recently, \cite{Abaurrea2018}
considered both maximum and minimum temperatures.
\cite{Perkins13} and \cite{Smith13} addressed the issue of analyzing different measurements and definitions of this phenomenon.

To circumvent the lack of agreement on heat wave definition, we work with the concept of an extreme heat event (EHE); this enables us to capture the notion of a period of persistent extremely high temperatures.
EHEs are based on an exceedance of threshold approach, often used in the analysis of extremes \citep{Davison1990}.
Then, at a given location, an EHE is defined as a run of consecutive daily temperature observations exceeding the threshold for that location.
Useful thresholds should be based on local conditions; we adopt the $95$th percentile of local daily maximum temperatures over a ten year period.

Much work on extreme events focuses on the analysis of their occurrence using asymptotic results from extreme value theory.
These results state that the occurrence of peaks over threshold follow a Poisson process under mild conditions \citep{Ogata88, Garcia14, Abaurrea15}.
A limitation of this approach is that important behaviors describing the nature of an EHE, such as the duration, average exceedance, or maximum exceedance above the threshold, are not captured.

Models for heat waves and extreme heat events commonly consider the occurrence of the events at particular locations, which are extracted from a time series of temperatures over a window of time.
However, spatial aspects of this phenomenon are also of interest and should be introduced into the modeling process, particularly since there is interest in predicting EHE behavior to locations without time series of temperatures.
Since spatial dependence is anticipated in the occurrence of EHEs, a joint model for time series at different locations that incorporates spatial dependence will borrow strength from nearby time series resulting in an expected increase in the probability of an EHE occurring at locations near where an EHE did occur.
Given the evidence of changing climate, the proposed models should also be able to capture nonstationary behavior in maximum daily temperature in time.
This can be done by allowing temporal random effects, time varying coefficients, or time-varying covariates \citep{Cheng14, Abaurrea15}.

The contribution of this paper is to take up the challenge of providing a methodology which incorporates spatial dependence to study the occurrence and behavior of extreme heat events during a period of time over a specified region. The intent is to enable more complete characterization of the events.
We develop a space-time model based on a point-referenced collection of temperature time series that enables the prediction of both the incidence and the characteristics of the EHEs occurring at any location in the region.

A direct approach considers a spatial model for daily maximum temperatures which can then be used to characterize the EHEs. A shortcoming of this approach is that such a model will be driven by the bulk of the distribution, i.e., where most of the data is observed. The model can yield poor fits for the upper tail where the main interest for the model lies when attempting to characterize EHEs \citep{Keellings15, Shaby16}.
A more promising approach to effectively model both the extremes as well as the bulk of the distribution is to consider a model incorporating thresholding. Here, we propose a model which switches between two observed states, one that defines extreme heat days (those above the temperature threshold) and the other that defines non-extreme heat days (those below the temperature threshold).
Importantly, this two-state structure allows temporal dependence of the observations but also that the parameters controlling the effects of covariates and the spatial dependence can differ between the two states.

The resulting model provides a mixture distribution \citep[in the spirit of][]{Behrens2004, Macdonald2011, Scarrott2016} at locations and times. It employs truncated multivariate t-distributions for the upper tails to capture tail dependence over time and space, with a truncated multivariate normal distribution for the bulk of the data which is below the threshold.  At a given location, the mixture weights for each day arise as the daily transition probabilities between the two observed states driven by a two-state autoregressive Markovian switching specification.
For daily maximum temperatures, we are able to demonstrate substantially improved out-of-sample predictive performance compared with a corresponding spatial autoregressive model, using the multivariate t specification that ignores thresholds.

The full model is specified in a hierarchical Bayesian framework and estimated using a Markov chain Monte Carlo (MCMC) algorithm. In this framework, posterior predictive distributions of the foregoing features of the EHEs (rate of incidence, duration, maximum exceedance, average exceedance) can be easily obtained. Specifically, for a given site,
we can obtain posterior predictive samples of the time series of daily maximum temperatures from which posterior predictive samples of the characteristics of their EHEs can be extracted.

There is limited literature using Markov models to analyze EHEs, however none of this modeling introduces spatial structure.
\cite{Smith97} studied daily temperature exceedances over a threshold and fitted a Markov model to them. \cite{Shaby16} proposed a two-state switching hidden Markov model and used a latent variable that controlled whether a day was assigned to the heat wave or the non-heat wave state. By contrast, given local thresholds, our approach takes these state variables as observed rather than latent. Spatial dependence in extreme temperatures has been recognized in the literature \citep[see, e.g.,][]{Davison2012, Shaby2012, Fuentes2013, Thibaud2016} but using different approaches.
For instance, \cite{Cooley07} proposed a Bayesian spatial modeling of extreme precipitation in order to obtain maps of precipitation return levels. \cite{Reich19} suggested a spatial Markov model for climate extremes, using latent clustering of neighboring regions. \cite{Guillot15} used Markov random fields to model high-dimensional spatial fields for paleoclimate reconstructions.

There are several motivations for building an effective space-time model.  The model can be used to make predictions of EHE characteristics at unobserved locations.
These predictions are important for understanding the possible impacts of EHEs on human health, ecosystems, and the economy, since the assessment of these effects may require predictions at locations where monitoring devices are not available.
Such spatial prediction may also be useful to complete observed series with missing information.
A further potential application of the model is to make predictions of the spatial extent of an EHE, defined as the geographical area (often expressed as a percentage of the region) affected by extreme temperatures at a given time.
Lastly, the model can be used to provide information for a deeper understanding regarding the temporal evolution of aspects related to temperatures across a region. For instance, it can be used to study changes in the incidence of EHEs and their characteristics between decades, to assess whether the factors affecting the non-extreme and the extreme daily maximum temperatures are the same, or  to study EHE behavior averaged over a region. Spatial uncertainty with regard to all of these concerns is provided.

As a final comment, there is a useful distinction between the objective of modeling daily maximum temperatures and learning about characteristics of extreme heat events. The distinction is analogous to that of predicting weather versus climate. That is, like weather, if the goal is to predict an annual time series of daily maximum temperatures, our autoregressive two-state threshold based model will not outperform a corresponding autoregressive model which ignores thresholds. However, like climate, if our goal is to predict incidence and characteristics of EHEs over a window of time, say, a decade, then our two-state model will consequentially outperform the corresponding autoregressive model.


The outline of the paper is as follows. 
Section \ref{Sec:Data} presents an exploratory analysis of the dataset consisting of approximately 60 years of daily maximum temperatures at 18 locations throughout the Comunidad Aut\'onoma de Arag\'on in Spain.
Section \ref{Sec:Model} describes the modeling details.
Comparison of the proposed two-state model to a corresponding autoregressive model ignoring the threshold using leave-one-out validation is presented in Section \ref{Sec:Fit}.
Results and some associated inference summaries are supplied in Section \ref{Sec:Results}.
Finally, Section \ref{Sec:Disc} concludes with a summary and possible directions for future work.

\section{Exploratory data analysis}
\label{Sec:Data}

The region of interest is the Comunidad Aut\'onoma de Arag\'on region in northeastern Spain, located in the Ebro basin (85,362 km$^2$).
The Ebro river flows from the NW to the SE through a valley bordered by the Pyrenees and the Cantabrian Range in the north and the Iberian System in the southwest. The maximum elevation is approximately 3400 m in the Pyrenees, 2600 m in the Cantabrian Range and Iberian System, and between 200-400 m in the central valley. In general, the area is characterized by a Mediterranean-continental dry climate with irregular rainfall, and a large temperature range.
However, several climate subareas can be distinguished due to the heterogeneous orography and other influences, such as the Mediterranean sea to the east, and the continental conditions of the Iberian central plateau in the southwest. Zaragoza, the largest city in the region, is located in the central part of the valley, and experiences more extreme temperatures and drier conditions.
This variation in climate conditions suggests that Arag\'{o}n is an interesting study region.

Our data are observational series from AEMET (the Spanish Meteorological Office). 
Only long term series with limited missing observations were considered, resulting in daily maximum temperatures for 18 sites 
across and around the Comunidad Aut\'onoma de Arag\'on region for the years 1953-2015. 
We note that series from other sources, e.g., satellite data or reanalysis, would not be compatible with this observational data.

\begin{figure}
\begin{center}
\includegraphics[scale=.5]{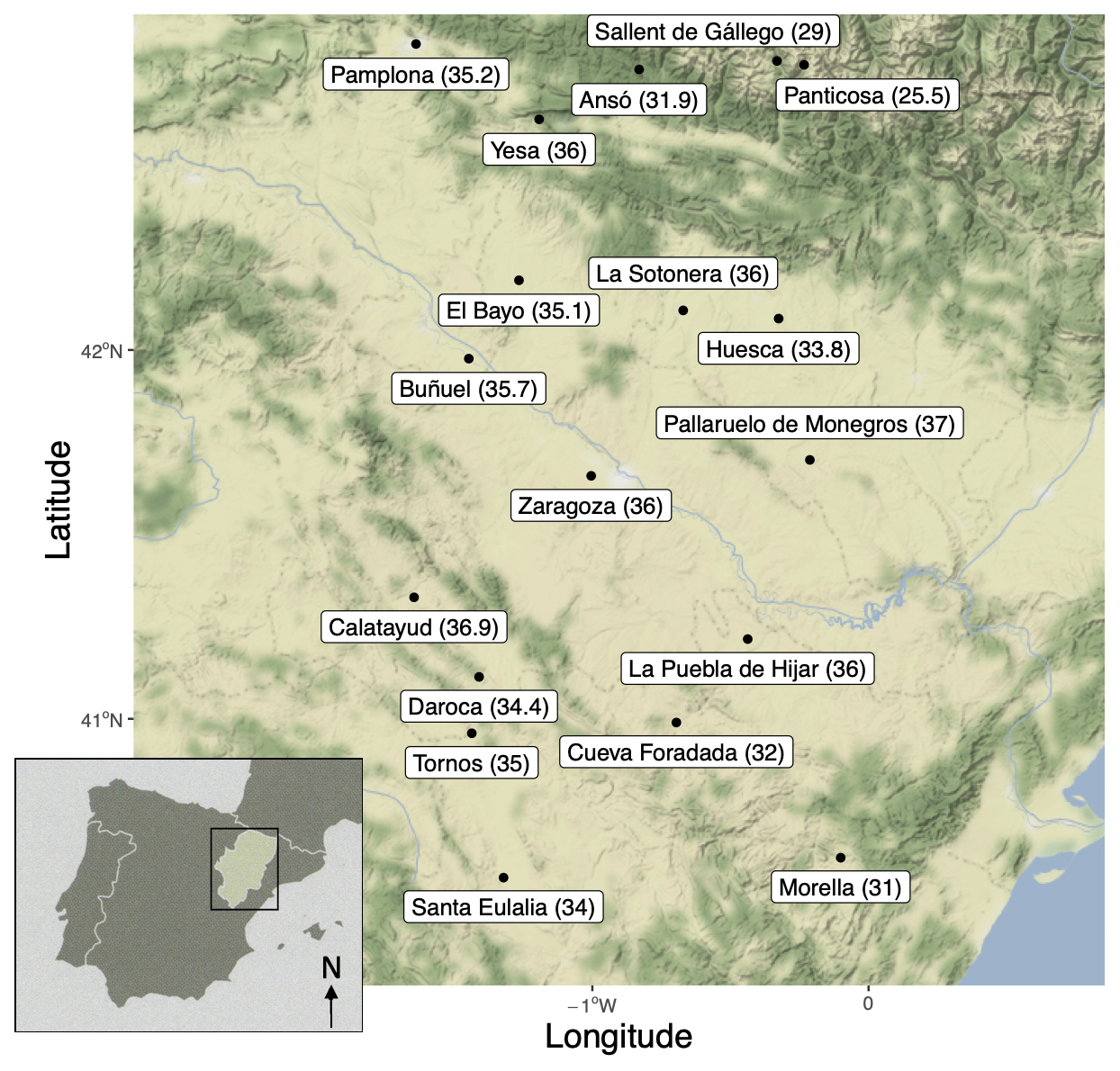}
\caption{Spatial map of the 18 sites across and around the Comunidad Aut\'onoma de Arag\'on region in northeastern Spain used in the analysis with associated 95th percentile (threshold) of daily maximum temperature for the months June, July, and August during 1953-1962. \label{fig:Map}}
\end{center}
\end{figure}

\begin{figure}
\begin{center}
\includegraphics[scale=.5]{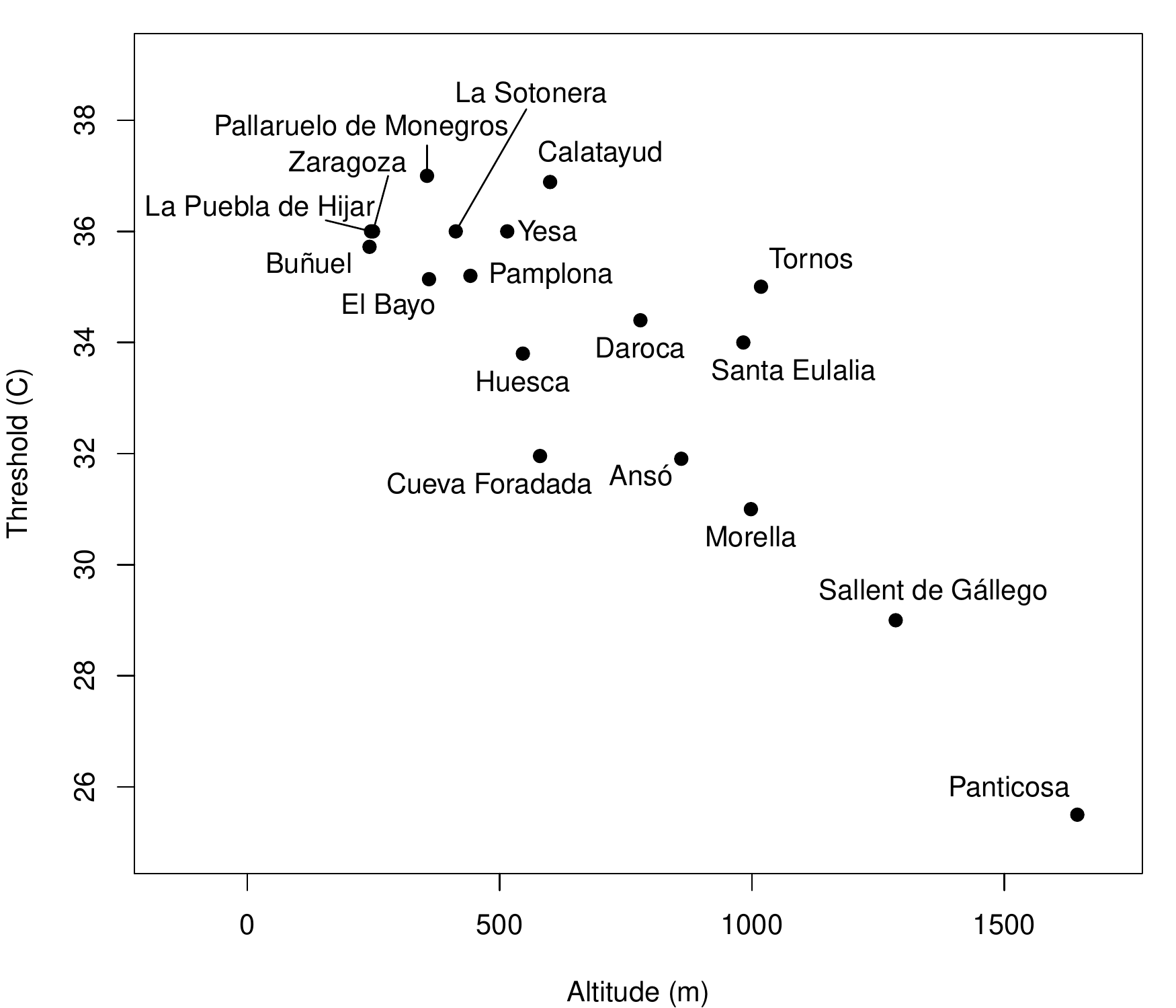}
\caption{The temperature threshold versus altitude (meters) for the 18 sites. \label{fig:Q}}
\end{center}
\end{figure}

The names and locations of the 18 sites are shown spatially in Figure \ref{fig:Map}.
Data from the years 1953-1962 were used to obtain the location-specific thresholds for extreme heat events and the data for the years 1963-2015 were used in the modeling.
Specifically, the 95th percentile of daily maximum temperature for the months June, July, and August during the 10 years 1953-1962 was computed for each site to determine the \emph{threshold} for defining extreme heat events at the site. 
This period, being the oldest in the dataset, seems most appropriate to use for obtaining thresholds in order to investigate potential change in EHE behavior over time.
The 95th percentiles are reported in parentheses in Figure \ref{fig:Map} and shown versus elevation/altitude for each site in Figure \ref{fig:Q}.
In general, there is a negative, roughly linear relationship between altitude and daily maximum temperature percentile.
In addition, while higher elevations are found in both the northern and southern part of the Arag\'{o}n region, the 95th percentiles tend to be lower in the north than south indicating a latitudinal gradient in addition to an elevation gradient for daily maximum temperature.

Monthly averages of daily maximum temperature for each location during for the years 1976-1985 and 2006-2015 are supplied in Figure \ref{fig:TS} of the supplementary material. Also shown are the differences (later minus earlier) in these monthly averages between the two time periods for the months April through October.  We see an increase in April and very sharp increase in May across all sites, indicating that warmer weather is now starting earlier.  We continue to see an increase through the summer months but in September we see a decrease for many sites suggesting an earlier end to summer at these sites.
Raw summaries of the daily maximum temperature data are included in Tables \ref{Tab:DaysAbove} and \ref{Tab:Events}.
The total number of days with maximum temperature above the specified threshold for two 10 year periods (1976-1985 and 2006-2015) are given in Table \ref{Tab:DaysAbove}.
These summaries are computed for the months June, July, and August.
In general, July and August have the greatest number of days with maximum temperatures above the threshold and the latter time period experienced more days above the threshold.
Table \ref{Tab:Events} shows the number of unique heat events, which considers each run of consecutive days having a maximum temperature above the threshold as one event.
Thus, the number of distinct extreme heat events is less than the number of days above the threshold since many events last more than one day.
In Table \ref{Tab:Events} each event is allocated to the month in which the event starts.
Again, the majority of heat events occur in July and August for each of the locations and more events occurred in the latter decadal period.

\begin{table}[ht]
\caption{The cumulative number of days with maximum temperature above the threshold for each location for the months June, July, and August for the two ten year windows, 1976-1985 and 2006-2015. \label{Tab:DaysAbove}}
\centering
\begin{tabular}{lrr|rr|rr}
 \hline
 & \multicolumn{2}{c}{June} & \multicolumn{2}{c}{July} & \multicolumn{2}{c}{August}  \\
 Location & '76-'85 & '06-'15& '76-'85 & '06-'15& '76-'85 & '06-'15\\
 \hline
 Pamplona & 0 & 7 & 19 & 26 & 1 & 32  \\
 Bu\~{n}uel & 7 & 27 & 39 & 74 & 9 & 45 \\
 El Bayo & 11 & 15 & 43 & 42 & 23 & 36\\
 Morella & 2 & 14 & 52 & 48 & 26 & 46  \\
Huesca & 16 & 32 & 72 & 112 & 35 & 79  \\
 Tornos & 5 & 13 & 40 & 55 & 18 & 50  \\
 Santa Eulalia & 12 & 10 & 71 & 57 & 36 & 35  \\
 Calatayud & 0 & 9 & 20 & 20 & 7 & 30  \\
 Panticosa & 4 & 13 & 45 & 60 & 16 & 41  \\
 Puebla de H\'{\i}jar & 23 & 26 & 87 & 82 & 49 & 62  \\
 Ans\'{o} & 7 & 9 & 72 & 39 & 44 & 43  \\
  Daroca & 6 & 20 & 63 & 92 & 28 & 76 \\
 Zaragoza & 6 & 26 & 42 & 89 & 13 & 57  \\
 La Sotonera & 6 & 11 & 34 & 41 & 17 & 38 \\
 Pallaruelo de Monegros & 7 & 12 & 33 & 36 & 16 & 26  \\
Cueva Foradada & 12 & 41 & 89 & 124 & 39 & 96  \\
Sallent de G\'{a}llego & 11 & 15 & 85 & 47 & 44 & 47  \\
 Yesa & 6 & 7 & 45 & 22 & 24 & 28  \\
  \hline
\end{tabular}
\end{table}

\begin{table}[ht]
\caption{Number of unique extreme heat events by month and location for the two ten-year windows, 1976-1985 and 2006-2015. \label{Tab:Events}}
\centering
\begin{tabular}{lcc|cc|cc}
 \hline
 & \multicolumn{2}{c}{June} & \multicolumn{2}{c}{July} & \multicolumn{2}{c}{August} \\
  Location & '76-'85 & '06-'15& '76-'85 & '06-'15& '76-'85 & '06-'15\\
 \hline
 Pamplona & 0 & 4 & 8 & 20 & 1 & 17  \\
 Bu\~{n}uel & 6 & 13 & 21 & 35 & 8 & 22  \\
 El Bayo & 5 & 8 & 19 & 22 & 16 & 15  \\
 Morella & 2 & 8 & 23 & 21 & 14 & 22  \\
 Huesca & 6 & 13 & 30 & 47 & 19 & 27  \\
 Tornos & 3 & 5 & 20 & 32 & 10 & 23  \\
 Santa Eulalia & 5 & 5 & 30 & 24 & 13 & 15  \\
 Calatayud & 0 & 6 & 12 & 10 & 5 & 15  \\
 Panticosa & 2 & 5 & 20 & 27 & 10 & 17 \\
 Puebla de H\'{\i}jar & 12 & 11 & 32 & 37 & 25 & 22  \\
 Ans\'{o} & 3 & 4 & 26 & 23 & 22 & 19 \\
  Daroca & 4 & 10 & 24 & 41 & 13 & 32  \\
Zaragoza & 5 & 10 & 20 & 46 & 10 & 22  \\
 La Sotonera & 1 & 6 & 16 & 21 & 10 & 13  \\
 Pallaruelo de Monegros & 3 & 7 & 15 & 16 & 10 & 13  \\
 Cueva Foradada & 8 & 21 & 33 & 53 & 17 & 33 \\
 Sallent de G\'{a}llego& 4 & 5 & 24 & 20 & 20 & 21  \\
 Yesa & 4 & 3 & 22 & 13 & 15 & 12  \\
  \hline
\end{tabular}
\end{table}

Next, we investigate the autoregressive behavior of daily maximum temperature.
Let $Y_t(\bs)$ denote the maximum temperature on day $t$ at location $\bs$ and let $q(\bs)$ denote the threshold for location $\bs$ such that when $Y_t(\bs) \geq q(\bs)$ we have a day with extreme heat.
Additionally, define the binary indicator $U_{t}(\bs)$, where $U_{t}(\bs) =1$ indicates a daily maximum temperature above the threshold on day $t$ at location $\bs$, otherwise $U_{t}(\bs) =0$.

\begin{figure}
\begin{center}
\includegraphics[scale=.7]{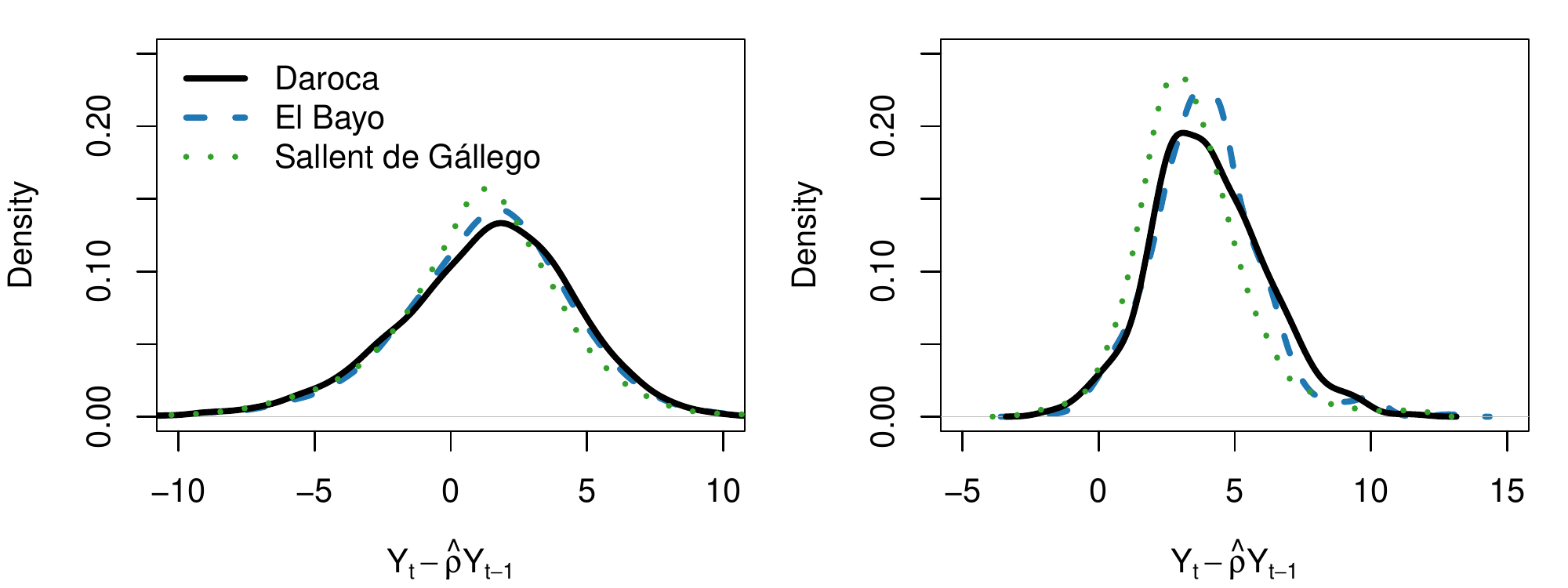}
\caption{Density estimates of (left) $(Y_{t}(\bs) - \hat{\rho}Y_{t-1}(\bs)|U_t(\bs)=0)$ and (right) $(Y_{t}(\bs) - \hat{\rho}Y_{t-1}(\bs)|U_t(\bs)=1)$ for three locations. \label{fig:AR}}
\end{center}
\end{figure}

In an exploratory mode, we fit independent AR(1) models $Y_t(\bs) = \rho(\bs) Y_{t-1}(\bs) + \epsilon_t(\bs)$ using the data from the years 1963-2012 for each location where $\rho(\bs)$ is the autoregressive parameter and $\epsilon_t(\bs)$ is pure error.
From the estimated $\hat{\rho}(\bs)$, we compute the kernel density estimate of $(Y_t(\bs) - \hat{\rho}(\bs) Y_{t-1}(\bs)|U_t(\bs))$ for both $U_t(\bs) = 0$ and $U_t(\bs) =1$.
Figure \ref{fig:AR} shows kernel density estimates for three representative locations. It reveals both the differences across locations as well as differences when conditioning on $U_t(\bs)$.
Notably, the distribution $(Y_t(\bs) - \hat{\rho}(\bs) Y_{t-1}(\bs)|U_t(\bs)=0)$ is much more diffuse than that of $(Y_t(\bs) - \hat{\rho}(\bs) Y_{t-1}(\bs)|U_t(\bs)=1)$ for all three locations.

Finally, we explore the variability in temperature for days above threshold and days below threshold by location.
For each location, we computed the variance in maximum daily temperature for the days above threshold and days below threshold, as well as the ratio of these variances.
As expected, there is much more variation in maximum daily temperatures for temperatures below the threshold than above threshold.
The location-specific variances in daily maximum temperature for the below threshold days range from 49.6 to 83.2 with a mean of 65.0$(^{\text{o}}$C)$^{2}$, whereas the above threshold days range from 1.17 to 2.45 with a mean of 1.72$(^{\text{o}}$C)$^{2}$.  
Lastly, we conducted exploratory data analysis to investigate possible changes in variability over time.
We compared these location-specific variance estimates between the two ten-year time windows and did not detect any significant differences.

\section{Modeling details}
\label{Sec:Model}
Let $Y_t(\mathbf{s})$ denote the daily maximum temperature at time $t$ and location $\mathbf{s}$.
We propose a two-state model where the state for a given day defines whether the location is experiencing an extreme heat event or not.
Let $U_t(\mathbf{s}) \in \{0,1\}$ denote the state at time $t$ for location $\mathbf{s}$ where a value of 0 denotes the below threshold state and a 1 denotes the above threshold state. We specify a threshold, $q(\bs)$ at each observed location, developed as discussed in Section 2. Again, we are interested in EHEs at $\bs$ relative to the climate at $\bs$.
Here, $U_{t}(\bs)$ is a spatial binary time series process reflecting times of transition or state-switching. It is observed for each $t$ at a monitored site but is latent elsewhere.

Under a typical hidden Markov model, we would assume $U_t(\mathbf{s})$ is a Markov process where the state $U_t(\mathbf{s})$ depends only on the previous state $U_{t-1}(\mathbf{s})$.
Then, the distribution of $Y_t(\mathbf{s})$ would be specified explicitly given $U_t(\mathbf{s})$. Here, we take a different approach since our states are not hidden.
That is, the states are observed given a threshold and the temperatures are modeled under the restriction to a given state.
Conversely, given the threshold, $U_{t}(\bs)$ is a binary function of $Y_{t}(\bs)$.
Furthermore, our model specification allows the transition probabilities between states to be a function of previous temperature.

\begin{figure}
\begin{center}
\includegraphics[scale=.28]{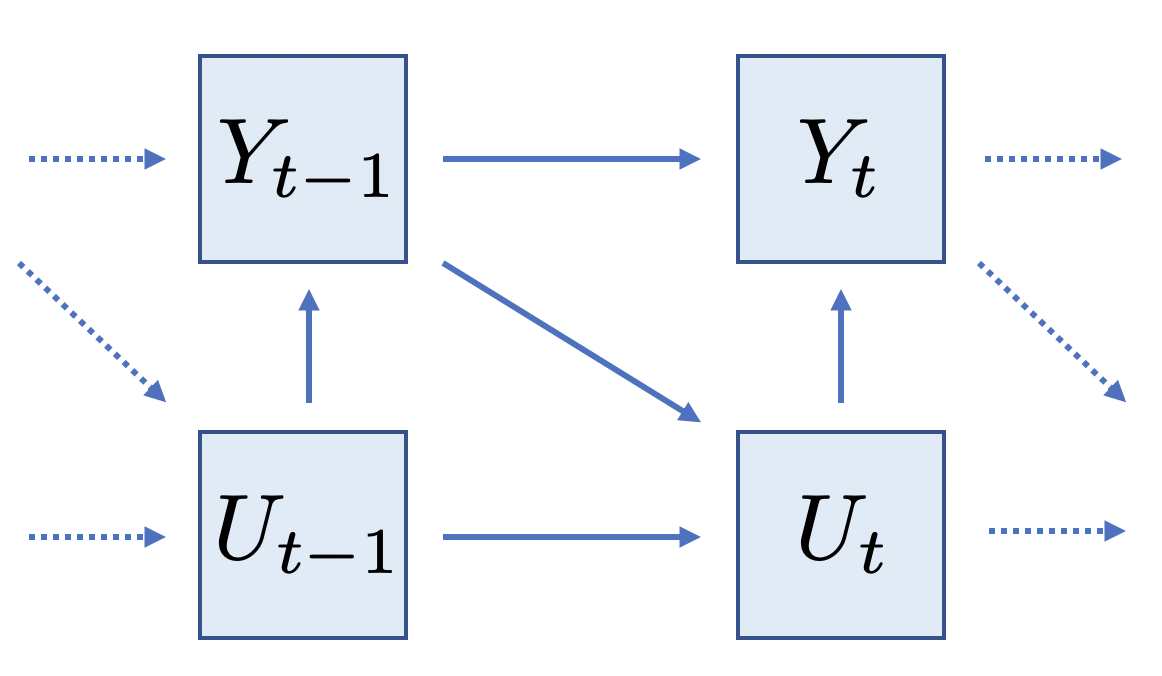}
\caption{Structural diagram showing the relationships between the binary process, $U_{t}(\bs)$, and temperature $Y_{t}(\bs)$ for a given location. \label{fig:Diag}}
\end{center}
\end{figure}

This specification ensures the transition probabilities to be ``local'', i.e., to vary with location and to depend upon the previous day's maximum temperature at that location.
For example, if the previous day's maximum temperature was sufficiently high so that we were in an extreme heat state, then we would expect this to affect the probability of staying in the extreme heat state for the next day. 
Since EHEs are fairly rare, we would expect a higher probability of being in an EHE state at day $t$ if we were in an EHE state at day $t-1$ than if we were not in an EHE state at day $t-1$. The opposite would be expected if the maximum temperature of the previous day resulted in a non-extreme heat state.  
This defines the notion of persistence of extreme heat, which is a main focus of our investigation. 

Two remarks are useful here. First, we could condition on a longer history of maximum temperatures or an average of previous maxima. We experimented with introducing additional lags in the modeling but found no gain in predictive performance. Second, failing to condition on the previous day's temperature provides transition probabilities which don't predict well the persistence of above or below threshold states.
Thus, we supply a location-specific autoregressive model for current daily maximum temperature which depends upon the current day state as well as the previous day's maximum temperature.

We define our joint distribution for temperature and state in a first order Markov fashion explicitly as follows.
Given $Y_{t-2}(\bs)$, and thus, $U_{t-2}(\bs)$, for two consecutive time points, $t-1$ and $t$, we write the joint distribution $[U_{t-1}(\mathbf{s}),Y_{t-1}(\mathbf{s}), U_{t}(\mathbf{s}), Y_{t}(\mathbf{s})]$ as

\begin{equation*}
[Y_{t}(\mathbf{s})|U_{t}(\bs), Y_{t-1}(\mathbf{s})][U_{t}(\mathbf{s})|Y_{t-1}(\mathbf{s})][Y_{t-1}(\mathbf{s})|U_{t-1}(\mathbf{s}), Y_{t-2}(\bs)][U_{t-1}(\mathbf{s})|Y_{t-2}(\bs)].
\end{equation*}
Hence, the process is started with $Y_{0}(\bs)$, followed by the distribution $[U_1(\bs)|Y_0(\bs)]$, and so on.
Figure \ref{fig:Diag} provides a simple graph of the specification and also reveals how it differs from a customary dynamic model specification.

This formulation requires three model specifications:
\begin{enumerate}[(i)]
\item $[Y_{t}(\mathbf{s})|U_{t}(\bs) = 0, Y_{t-1}(\mathbf{s})]$
\item $[Y_{t}(\mathbf{s})|U_{t}(\bs) = 1, Y_{t-1}(\mathbf{s})]$
\item $[U_{t}(\mathbf{s})|Y_{t-1}(\mathbf{s})]$.
\end{enumerate}
With $q(\bs)$ denoting the threshold (quantile) for location $\bs$,
we need truncated distributions for (i) and (ii), i.e., $[Y_{t}(\bs) = y]1( y < q(\bs))$ and $[Y_{t}(\bs) = y]1(y \geq q(\bs))$, respectively.
For (i), we adopt truncated normal distributions, with autoregressive centering,

\begin{equation}
TN\left(\mu_{t}^{0}(\bs)-\rho^{0}(Y_{t-1}(\bs)- \mu_{t-1}^{0}(\bs)), \sigma^{2,0}(\bs)\right) \text{I}(-\infty, q(\bs)).
\label{eq:TN}
\end{equation}
%
Details for $\mu_{t}^{0}(\bs)$ are given below.

When $U_{t}(\mathbf{s}) = 1$, we are examining upper tail behavior so we would seek to incorporate upper extremal dependence.  For a multivariate distribution, this leads to examination of the conditional probability of exceedance of a large threshold for some components given exceedance for some others \citep[see, e.g.,][]{Chan2008}.
With time series, as in our case, if $Y_{t-1} \sim G$ and $Y_{t} \sim H$ where $G$ and $H$ are cdfs, the bivariate extreme value coefficient can be written as $\gamma \equiv \text{lim}_{u \rightarrow 1^{-}}P(H(Y_{t}) > u| G(Y_{t-1}) >u)$.  We have tail dependence when $\gamma > 0$.  
The multivariate normal distribution is known to exhibit tail independence ($\gamma=0$).  However, \cite{Chan2008} develop expressions for multivariate  tail dependence for the multivariate t-family of distributions and show that they are nonzero for any finite degrees of freedom.  
While we are not concerned with calculation of these coefficients, we can incorporate tail dependence into our modeling by employing a truncated t-distribution for the case when $U_{t}(\mathbf{s})=1$. 
\cite{Schoelzel2008} also considered multivariate t distributions in the context of extremes in temperature and precipitation.

Specifically, for (ii), we adopt a truncated t distribution, with autoregressive centering,

\begin{equation}
Tt\left(\mu_{t}^{1}(\bs)-\rho^{1}(Y_{t-1}(\bs)- \mu_{t-1}^{1}(\bs)), \sigma^{2,1}(\bs)\right) \text{I}(q(\bs), \infty).
\label{eq:Tt}
\end{equation}
Details for $\mu_{t}^{1}(\bs)$ are given below.  As an aside, because we have multivariate t-distributions in both time and space, we immediately inherit tail dependence in space as well.

A common distribution for exceedances (i.e., the upper tail) would be the generalized Pareto distribution (GPD) \citep{Coles2001}, which enables a threshold, scale, and shape parameter.
The GPD is unsuitable in our setting since we are not modeling an autoregressive threshold. 
Rather, we are modeling autoregression in the
data, and therefore need an autoregressive specification in a centering
parameter.  
Autoregression in the scale parameter of the GPD is inappropriate since it misaligns  dependence on the data scale with dependence on the variability scale. 



Exploratory analysis in the previous section suggested a smaller variance for the above threshold daily maximum temperature distribution than for the below threshold daily maximum temperature distribution. This seems evident since the support for daily maximum temperatures above the threshold is much shorter than the support for daily maximum temperatures below the threshold. Further, we introduce spatially-varying variances, expecting that variation in say, Jaca (in the Pyrenees in the north of the region) would be different from variation in say, Zaragoza (flat and central in the region). 
This was also suggested by the exploratory data analysis in Section \ref{Sec:Data}.

For (iii) we introduce a probit link to define
\begin{equation}
\Phi^{-1}(p_{t}(\bs)) \equiv \Phi^{-1}(P(U_{t}(\bs)=1|Y_{t-1}(\bs))) \equiv \eta_{t}(\bs)
\label{eq:U}
\end{equation}
with $\eta_{t}(\bs)$ given below.  Putting (i), (ii), and (iii) together, we have created a mixture distribution for $Y_{t}(\bs)$.  We have a truncated normal distribution for the bulk of the distribution, a truncated t distribution for the upper tail of the distribution, and mixture weights according to $P(U_{t}(\bs) =0)$ or $P(U_{t}(\bs) =1)$, respectively. We note that this mixture distribution has a discontinuity in the density at $q(\bs)$ but, importantly, the cdf is continuous.  


Now, we turn to the specifics of $\mu_t(\bs)$ and $\eta_t(\bs)$. Compacting notation, for $\mu_{t}^{U_t(\bs)}(\bs)$ we consider
\begin{equation}
\begin{split}
\mu_{t}^{U_t(\bs)}(\bs) = \beta_0^{U_t(\bs)} + \beta_0^{U_t(\bs)}(\bs) + &\gamma_{[\frac{t}{365}]+1}^{U_t(\bs)} + \beta_{1}^{U_t(\bs)}\text{elev}(\bs) + \beta_{2}^{U_t(\bs)}\text{lat}(\bs)\\&+ \lambda_{1}\text{sin}(2\pi t/365) + \lambda_{2} \text{cos}(2\pi t/365).
\end{split}
\end{equation}
Here, $\beta_0^{U_t(\bs)}$ denotes a global (across the domain for our dataset) intercept and $\beta_0^{U_t(\bs)}(\bs)$ denotes a local spatial intercept, i.e., providing local adjustment to the global intercept. Each $\beta_0^{U_t(\bs)}(\bs)$ is modeled as a mean $0$ Gaussian process with exponential covariance function. For $\gamma_{[\frac{t}{365}]+1}^{(1)}$, where $[\hspace{.2cm}]$ denotes the greatest integer function, we are counting \emph{years} with this subscript and, as a result, the $\gamma$'s provide annual intercepts to allow for yearly shifts, i.e., for hotter or colder years.
The \emph{sin} and \emph{cos} terms are introduced to capture annual seasonality with their coefficients reflecting associated amplitudes.
In leap years, the specification is analogous, replacing 365 with 366 days.
This seasonality is critical to ensure that an annual daily maximum temperature trajectory over the course of a year at a location will provide sensible realizations.

We note that with circularity in time, this definition yields a discontinuity in $\mu_{t}^{U_t(\bs)}(\bs)$ from December 31 to January 1.\footnote{We could start the year at another day but, regardless, we will always experience a jump on that day.} However, since time is modeled discretely, we do not expect that this will be a concern with regard to EHE behavior.
Elev$(\bs)$ is the elevation at $\bs$ and Lat$(\bs)$ is the latitude.
We explored additional potential site level covariates, e.g., slope, aspect, distance from water (distance from the Ebro river and the Mediterranean sea) but found no improved predictive performance.
Finally, $\rho^{U_t(\bs)}$ provides a \emph{centered} AR(1) specification, bringing in the previous day's temperature, $Y_{t-1}(\bs)$.

For $\eta_{t}(\bs)$ we propose

\begin{equation}
\begin{split}
\eta_{t}(\bs) = \phi_{0} + \phi_{0}(\bs) + \phi_{1}(Y_{t-1}(\bs) - q(\bs)) + \phi_{2}((&Y_{t-1}(\bs) - q(\bs))1(Y_{t-1}(\bs) - q(\bs) \geq 0)) \\
&+ \phi_{3} \text{sin}(2\pi t/365) + \phi_{4} \text{cos}(2\pi t/365).
\end{split}
\end{equation}
Here, centering by the threshold yields more sensible transition probabilities. $q(\bs)$ can be moved over to the intercept term in order to provide a spatially varying offset. However, the inclusion of $\phi_{0}(\bs)$, modeled as a Gaussian process, allows for a richer spatially-varying intercept. Writing $\eta_{t}(\bs)$ in terms of $Y_{t-1}(\bs)$ being a deviation from $q(\bs)$ suggests that there is no need to model $\phi_1$ or $\phi_2$ as spatially varying coefficients. The term with coefficient $\phi_{2}$ provides a slope adjustment according whether we are previously in state $1$ or in state $0$. This adjustment ensures continuity in $\eta_{t}(\bs)$ (hence, $p_{t}(\bs)$) as a function of $Y_{t-1}(\bs)$.

These specifications are offered as an attempt to provide regressors that capture the critical features that drive daily maximum temperatures with transitions relative to a threshold state. Other variations of these specifications could be considered. For instance,
indicator functions could be considered to introduce an intercept and/or slope adjustment. Also interaction effects might be examined.
We explored a few of these richer model specifications and found no additional benefits in model performance.

\section{Model fitting and comparison}
\label{Sec:Fit}

We briefly summarize the complete model specification including priors as well as the model fitting and its challenges. Then, we turn to the out-of-sample model comparison.

\subsection{Full specification of the model and fitting details}

Model inference is obtained in a Bayesian framework, requiring prior distributions for each of the model parameters.
When possible, diffuse and conjugate prior distributions are assigned.
We start with the parameters of models (i) and (ii), $[Y_{t}(\mathbf{s})|U_{t}(\bs) = 0, Y_{t-1}(\mathbf{s})]$ and $[Y_{t}(\mathbf{s})|U_{t}(\bs) = 1, Y_{t-1}(\mathbf{s})]$.
Recall that model (i) is a truncated normal distribution and (ii) is a truncated t distribution, for which we assume 3 degrees of freedom, the smallest choice to provide existence of second moments.
The autoregressive parameters $\rho_{0}$ and $\rho_1$ are each assigned a non-informative and independent Uniform $(-1,1)$ prior distribution.
The coefficient parameters $\beta_0^{0}$, $\beta_1^{0}$, and $\beta_2^{0}$ as well as $\beta_0^{1}$, $\beta_1^{1}$, and $\beta_2^{1}$ are each assigned independent and diffuse normal prior distributions with mean 0 and standard deviation 100.
Independent normal prior distributions with mean 0 and standard deviation 100 are also assigned to the coefficients of the seasonal terms, $\lambda_1$ and $\lambda_2$.
The yearly random effects $\gamma_{\left[\frac{t}{365}\right]+1}^{0}$ and $\gamma_{\left[\frac{t}{365}\right]+1}^{1}$ are assigned normal distributions with mean 0 and standard deviation 1. For identifiability, each of the random effects for the first year are fixed to 0.

Mean 0 Gaussian process priors are assumed for both $\beta_{0}^{0}(\bs)$ and $\beta_{0}^{1}(\bs)$.
The spatial covariance matrix is specified using the exponential covariance function.
The variance parameter for each of the spatial covariances is assigned an independent Inverse-Gamma (2,2) prior distribution.
Similarly, the spatially varying variance parameters, log($\sigma_{0}^{2}(\bs))$ and log$(\sigma_{1}^{2}(\bs))$, are also assigned independent Gaussian process priors.
Hyperpriors are assigned to the mean and variance of both processes; each mean is assigned a N(0,1) and each variance is assigned an Inverse-Gamma(2,2).
Whereas the other Gaussian processes were mean 0, specifying a hyperprior for the mean of the spatial variances enabled Bayesian learning with respect to the variances of the above and below threshold processes.
An exponential covariance function is again used to specify the spatial dependence.

The following priors are assigned to the parameters of model (iii), $[U_{t}(\mathbf{s})|Y_{t-1}(\mathbf{s})]$.
Here, independent normal prior distributions with mean 0 and standard deviation 100 are assigned to the coefficients $\phi_0$, $\phi_1$, $\phi_2$, $\phi_3$, $\phi_4$
The spatial random effect $\phi_{0}(\bs)$ is assigned a mean 0 Gaussian process prior.
The variance is assigned an Inverse-Gamma (2,2) hyperprior and the spatial covariance is again specified with an exponential covariance function.

Altogether, the model consists of five spatial Gaussian processes, $\beta_{0}^{0}(\bs)$, $\beta_{0}^{1}(\bs)$, $\text{log}(\sigma_{0}^{2}(\bs))$, \allowbreak $\text{log}(\sigma_{1}^{2}(\bs))$, and $\phi_{0}(\bs)$.
Spatial models are needed for each of these components in order to be able to predict at new locations.
Each of these Gaussian processes is assumed independent with the range parameter of the exponential covariance function fixed such that the effective range is equal to 400 km. This distance is approximately the maximum distance across the Comunidad Aut\'onoma de Arag\'on region from the north to the south and was chosen to capture large-scale spatial dependence across the region. This choice assumes that the local behavior in daily maximum temperature will be captured by the autoregressive process in the model.

Markov chain Monte Carlo is used to obtain samples from the joint posterior distribution.
The sampling algorithm is a Metropolis-within-Gibbs algorithm.
Posterior draws of each of the spatial random effects are obtained using an efficient elliptical slice sampler \citep{Murray2010}.
Working with the multivariate t distribution brings a convenient model fitting benefit.  Since the multivariate t arises by a random scaling of a multivariate normal distribution, specifically, if $\mathbf{Z} \sim N(\bzero, \Sigma)$ then, if $\omega \sim IG(\nu/2, \nu/2)$, $T= \bmu + \sqrt{\omega}\mathbf{Z} \sim \text{multivariate}-t_{\nu}(\bmu, \Sigma)$.  Hence, in model fitting, we can work with a truncated normal distribution merely introducing an additional random $\omega$.  In the Bayesian framework using MCMC, the random scale parameter is included in the conditional normal distribution. Within the sampling algorithm, we iteratively update the model parameters in the specification for the upper tail using a truncated normal distribution with variance $\sqrt{\omega}\sigma^{2,1}(\mathbf{s})$ and then sample $\omega$ given the other parameters.

\subsection{Model comparison and inference}

For model comparison, we confine ourselves to just two models, the specification above and a corresponding specification which ignores state relative to threshold. This second model will be an AR(1) time series model for the daily maximum temperatures using t-distributed errors. This is analogous to our models in (i) above with $\mu_{t}(\bs)$.  We retain the upper tail dependence but now without the truncation according to the state. That is, we specify a t-distribution for $[Y_{t}(\mathbf{s})| Y_{t-1}(\mathbf{s})]$. In both models we adopt spatially-varying variances for prediction at new locations.
Under the simpler model, we impose thresholds on the posterior predictive distributions after model fitting in order to capture EHEs and their characteristics.

We acknowledge that this autoregressive t-process is not the most sophisticated single-state model we could attempt to develop for daily maximum temperature data and use for model comparison.  
However, our primary goal is to assess whether our autoregressive spatial two-state model out-performs the analogous single-state model with regard to learning about EHE behavior.

Model comparison is implemented by single-point deletion (leave-one-out) validation. While we have done single site deletions for all $18$ sites, we show results for three illustrative locations, one in the southern part of the region, one central, and one in the north.  With strong interest in capturing persistence of EHEs, comparison is made using conditional and marginal error rates (defined below) with regard to prediction of an EHE day.  We shall see that our proposed model is substantially better at such prediction.

Then, using our model, inference is provided, employing posterior predictive summaries, for the EHE characteristics presented in Section \ref{Sec:Intro} - duration, maximum exceedance above threshold and average exceedance above the threshold.  Such inference is provided for the entire time span of the data as well as by decade for two decades of interest.

The thresholds, $q(\bs)$, are known/observed for each monitoring site, adopting a $95$th percentile of the daily maxima at a site \citep[e.g.,][]{Abaurrea2007} as described in Section 2.  These values were used directly in the model fitting for the two-state models specified above and, therefore, are known for our leave-one-out validation.
For prediction beyond our $18$ sites, we can employ Spain02 \citep[][the updated version of Spain01]{Herrera2016}, available at http://www.meteo.unican.es/es/datasets/spain02. This dataset provides daily maximum and minimum temperatures, \emph{Tmax} and \emph{Tmin} from 1951 to 2015 in a $0.1^{\text{o}}$ (10km) regular grid. These series are obtained by numerical methods that yield a smoothed spatial temperature signal (especially in the tails). These series are not appropriate to model EHE but provide temperatures to develop suitable thresholds at any location using nearby grid points.


\section{Results of model comparison and inference}
\label{Sec:Results}
We begin with the results of the model comparison using leave-one-out cross validation and then provide the inferential summaries from our two-state model for the region of interest.
Comparison is between the two state model presented in Section 3 and the associated single state model above.
Each model was fitted to the daily maximum temperature data for the 50 years spanning 1966 to 2015.
Markov chain Monte Carlo was run for 200,000 iterations.
The first half of each chain was discarded as burn-in and the remaining samples were used for posterior inference and prediction.

\subsection{Model comparison}

A leave-one-out comparison of the model is carried out employing series from three sites, Zaragoza, Tornos and Yesa. 
For each of these locations in the hold-out set, the entire time series of daily maximum temperature is withheld during model fitting. 
These three time series show the variation in climate across the study region. Zaragoza is located in the center of the region, surrounded by other locations in the dataset with similar climate. Yesa is located in a valley in the northwest part of the region, with a climate that is quite different from that of the rest of locations in that area. Tornos is near the southwest border of the region with a climate much different from that of its neighbors to the east with the same latitude.

We conduct our model comparison through prediction of exceedance days which requires prediction of $U_t(\mathbf{s})$. Since $U_t(\mathbf{s})$ is a binary variable, it is natural to assess its performance in terms of misclassification error rates. There are two possible errors here: (i) predicting $U_t(\mathbf{s}) = 1$ when $U_{t;obs}(\mathbf{s}) = 0$ and (ii) predicting $U_t(\mathbf{s}) = 0$ when $U_{t;obs}(\mathbf{s}) = 1$. Arguably, the second error has more impact, since it means failing to predict well an exceedance event given that it happened.  Recall from (2) and (4), that our modeling for $U_{t}(\bs)$ is conditioned on $Y_{t-1}(\bs)$, not only on $U_{t-1}(\bs)$.
Therefore, while these locations are withheld during model fitting, predictions of exceedance days requires the observation of the previous day's maximum temperature for each out-of-sample location. 
Thus, these predictions and errors are computed using a one-day-ahead prediction scheme and the previous days observed daily maximum temperature. 


For days \emph{only} when $U_{t;obs}(\mathbf{s}) = 1$, three different measures of predictive error are considered, two conditional and one marginal error. The first conditional error is

\begin{equation}
\begin{split}
1-p_{t}^{(1)}(\mathbf{s}) &\equiv 
1-P\left(U_t(\mathbf{s}) = 1 | Y_{t-1}(\mathbf{s}) =y_{t-1,obs}(\bs), y_{t-1,obs}(\bs) \geq q(\mathbf{s})\right)\\
&= P(U_t(\mathbf{s}) = 0 | Y_{t-1}(\mathbf{s})=y_{t-1,obs}(\bs), y_{t-1,obs}(\bs) \geq q(\mathbf{s})).
\end{split}
\label{eq:cond}
\end{equation}
In \eqref{eq:cond}, we condition on an observed previous day maximum temperature which was an exceedance day. It gives the error in persistence, or rather, that of an additional extreme heat day given the previous day's extreme heat temperature, so that smaller errors of this type imply better prediction of persistence. The second conditional error is $1-p_{t}^{(0)}(\mathbf{s}) \equiv P(U_t(\mathbf{s}) = 0 | Y_{t-1}(\mathbf{s})=y_{t-1,obs}(\bs), y_{t-1,obs}(\bs) < q(\mathbf{s}))$, where we condition on a previous day's maximum temperature which was not an exceedance day; it gives the error in predicting the onset of an EHE.
The marginal error is  $1-p_t(\mathbf{s})=P(U_t(\mathbf{s})= 0 | Y_{t-1}(\mathbf{s})= y_{t-1,obs}(\bs))$, i.e., given the previous day's maximum temperature regardless of whether or not it was an exceedance.

Each error rate is estimated as the average of daily point estimates for the set of $U_{t;obs}(\mathbf{s}) = 1$ over a selected time window. 
While we can compute these error rates for any time window within the calendar year, we choose the 92 day window arising from the months June, July and August (JJA), since this is when most of the exceedance events occur. We can then average over a year, decade, or the entire time window.

\begin{figure}
	\begin{center}
		\includegraphics[scale=.68]{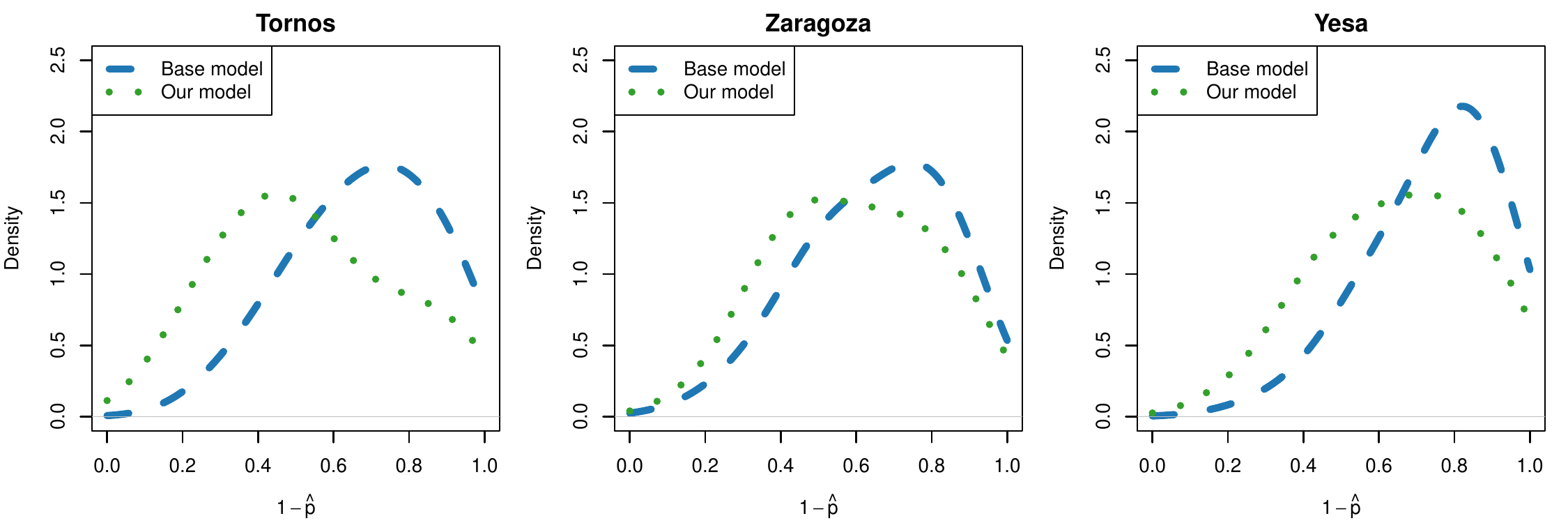}
		\includegraphics[scale=.68]{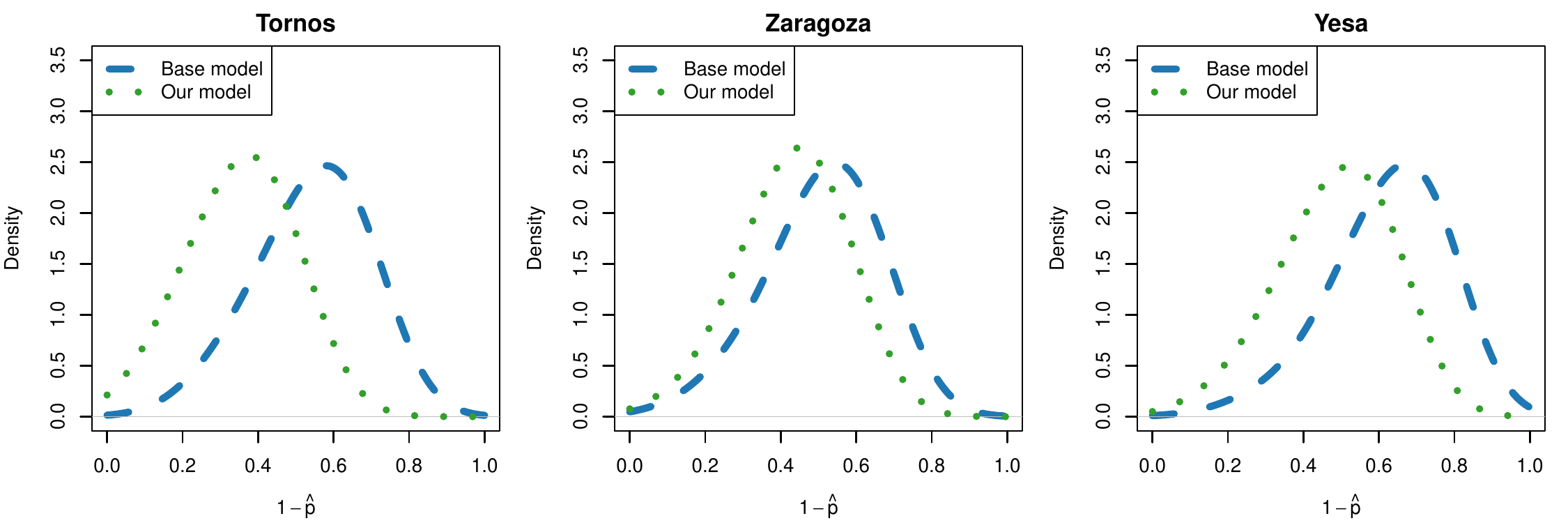}
		\caption{Distribution of $1-\hat{p}_t(\mathbf{s})$(top) and $1-\hat{p}_{t}^{(1)}(\mathbf{s})$ (bottom) calculated during JJA across all years for the out-of-sample locations.}
		\label{Fpt}
	\end{center}
\end{figure}

Figure \ref{Fpt} shows the distributions of $1-\hat{p}_t(\mathbf{s})$ and $1- \hat p_{t}^{(1)}(\mathbf{s})$, for all exceedance days in JJA across all years for the three out-of-sample locations. It is observed that in both cases, the distribution of the errors of the single-state model is shifted towards higher values, and the modes of the errors from the two-state model are lower than those from the single-state model.

The posterior predictive mean estimates of the error rates for all exceedance days in both models are summarized in Table \ref{TmeanErrors} for the two 10-year periods. 
The performance of the two-state model is clearly better, since the marginal errors and the errors conditioned on $y_{t-1,obs}(\mathbf{s}) \geq q(\mathbf{s})$ are lower in all the cases.
Therefore, the two-state model captures the observed persistence of an additional exceedance day better than the single-state model.
With regard to the conditional measures of error given $y_{t-1,obs}(\mathbf{s}) < q(\mathbf{s})$, those concerning prediction of the first day of an EHE, the two-state model outperforms the single-state model with less distinction. We see some gain in Tornos while in Zaragoza the errors are similar for both models.
While these error rates might seem large, they are computed using only exceedance days and have important interpretation and utility in terms of persistence.
Recall that exceedance days are not common since the location-specific thresholds are at the 95\% percentile. 
Further, when exceedance events occur, they are usually short in duration meaning their persistence is low. 
The error rates would be smaller if computed using all days in June, July, and August. 


To further investigate the predictive performance of the two-state model in terms of persistence, we expand our conditional measures of error.
Table \ref{TmeanErrors2} shows the posterior predictive mean estimates of the conditional error rates for all exceedance days, given that the observed previous daily maximum temperatures revealed an EHE had already persisted for one, two, or three days.
In general, the conditional error estimates are lower for those events with longer extents of persistence.
Compared to the single-state model, the error rates in the two-state model are lower for each location and extent of persistence.

\begin{table}[ht]
	\caption{Posterior predictive mean estimates of the error rates for all exceedance days, $1-\hat{p}_t(\mathbf{s})$, $1-\hat{p}_{t}^{(1)}(\mathbf{s})$ and $1-\hat{p}_{t}^{(0)}(\mathbf{s})$, in JJA for the periods 1976-1985 and 2005-2015 for the three out-of-sample locations.}
	\label{TmeanErrors}
	\centering
	\begin{tabular}{lrrrr}
		\hline
		&\multicolumn{2}{c}{1976-1985}&\multicolumn{2}{c}{2006-2015}\\
		\hline
		& Base & Our & Base & Our\\
		& Model & Model & Model & Model\\
		\hline
		Marginal Error:\\
		~~~~Tornos 	& 0.72 & 0.54 & 0.72 & 0.59 \\
		~~~~ Zaragoza & 0.70 & 0.62 & 0.64 & 0.59 \\
		~~~~ Yesa 	& 0.77 & 0.67 & 0.75 & 0.68 \\
		\hline
		Conditional Error, $y_{t-1,obs}(\mathbf{s}) \geq q(\mathbf{s})$:\\
		~~~~Tornos 	& 0.58 & 0.38 & 0.56 & 0.39 \\
		~~~~Zaragoza 	& 0.61 & 0.43 & 0.48 & 0.41 \\
		~~~~Yesa 	& 0.63 & 0.48 & 0.62 & 0.52 \\
		\hline
		Conditional Error, $y_{t-1,obs}(\mathbf{s}) < q(\mathbf{s})$:\\
		~~~~Tornos 	& 0.84 & 0.70 & 0.86 & 0.77 \\
		~~~~Zaragoza 	& 0.79 & 0.78 & 0.82 & 0.79 \\
		~~~~Yesa 	& 0.84 & 0.84 & 0.88 & 0.84 \\
		\hline
	\end{tabular}
\end{table}

\begin{table}[ht]
	\caption{Posterior predictive mean estimates of the error rates for all exceedance days, computed as $1-\hat{p}_t(\mathbf{s})$, in JJA for the whole period 1966-2015 for the three out-of-sample locations. The error rates are computed conditional on previous daily maximum temperatures such that the extreme heat event has already persisted for one, two, or three days.}
	\label{TmeanErrors2}
	\centering
	\begin{tabular}{lrr}
		\hline
		& Base & Our\\
		Conditional Error & Model & Model \\
		\hline
		$y_{t-1,obs}(\mathbf{s}) \geq q(\mathbf{s}), y_{t-2,obs}(\mathbf{s}) < q(\mathbf{s})$:\\
		~~~~Tornos 	& 0.59 & 0.39 \\
		~~~~Zaragoza 	& 0.64 & 0.48 \\
		~~~~Yesa 	& 0.61 & 0.54 \\
		$y_{t-1,obs}(\mathbf{s}) \geq q(\mathbf{s}), y_{t-2,obs}(\mathbf{s}) \geq q(\mathbf{s}), y_{t-3,obs}(\mathbf{s}) < q(\mathbf{s})$:\\
		~~~~Tornos 	& 0.52 & 0.34 \\
		~~~~Zaragoza 	& 0.58 & 0.41 \\
		~~~~Yesa 	& 0.56 & 0.44 \\
		$y_{t-1,obs}(\mathbf{s}) \geq q(\mathbf{s}), y_{t-2,obs}(\mathbf{s})\geq q(\mathbf{s}), y_{t-3,obs}(\mathbf{s}) \geq q(\mathbf{s}), y_{t-4,obs}(\mathbf{s}) < q(\mathbf{s})$:\\
		~~~~Tornos 	& 0.51 & 0.33 \\
		~~~~Zaragoza 	& 0.56 & 0.36 \\
		~~~~Yesa 	& 0.55 & 0.42 \\
		\hline
	\end{tabular}
\end{table}

 \subsection{Model adequacy}

We briefly consider model adequacy with regard to the main objective of our study: out-of-sample prediction of characteristics of extreme heat events.  We do this by comparing the posterior predictive distribution of exceedance days as well as EHE characteristics (duration, and intensity) with the observed empirical counterparts. That is, we generate entire posterior predictive time series for each of the hold-out sites and extract features of interest from each. Comparisons are made for the entire time window of the analysis, 1966-2015, as well as for the two 10-year periods, 1976-1985 and 2006-2015, to examine the time evolution.

\begin{figure}
\begin{center}
\includegraphics[scale=.68]{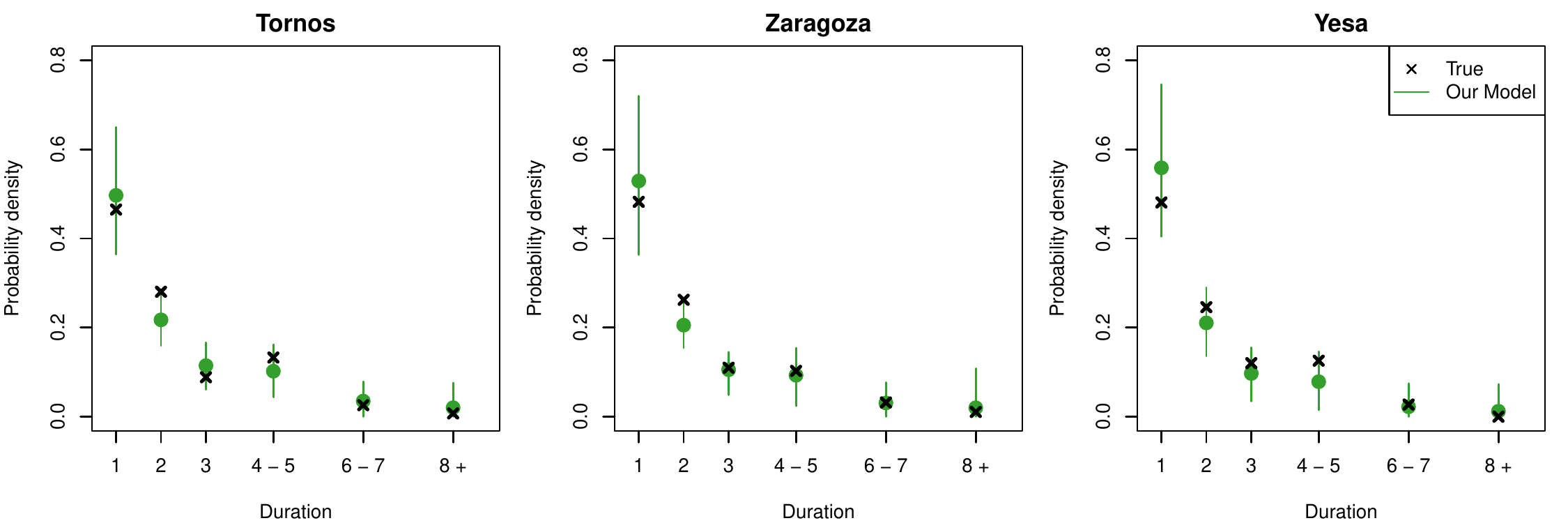}
\caption{Posterior predictive mean estimates and 90\% credible intervals of the probability density for the durations of extreme heat events across the years 1966-2015. For each of the three out-of-sample locations, true duration density is plotted for the durations 1, 2, and 3 days, 4-5 days, 6-7 days, and 8 or more days. \label{fig:DurationDist}}
\end{center}
\end{figure}

\begin{figure}
\begin{center}
\includegraphics[scale=.68]{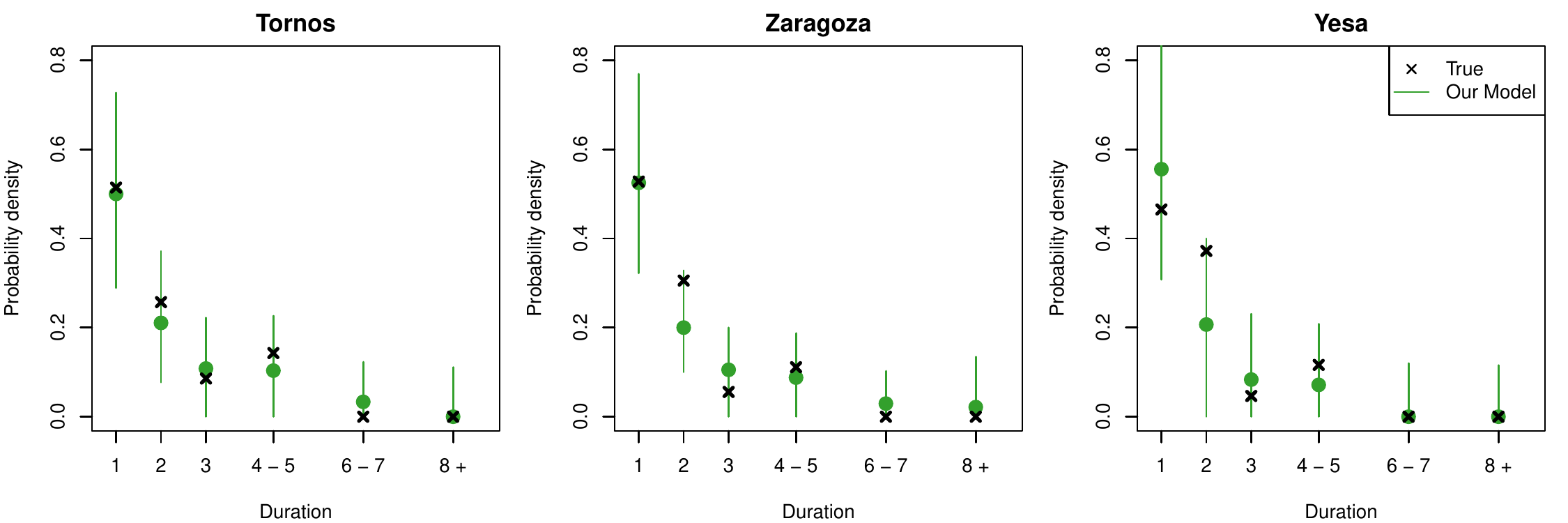}
\includegraphics[scale=.68]{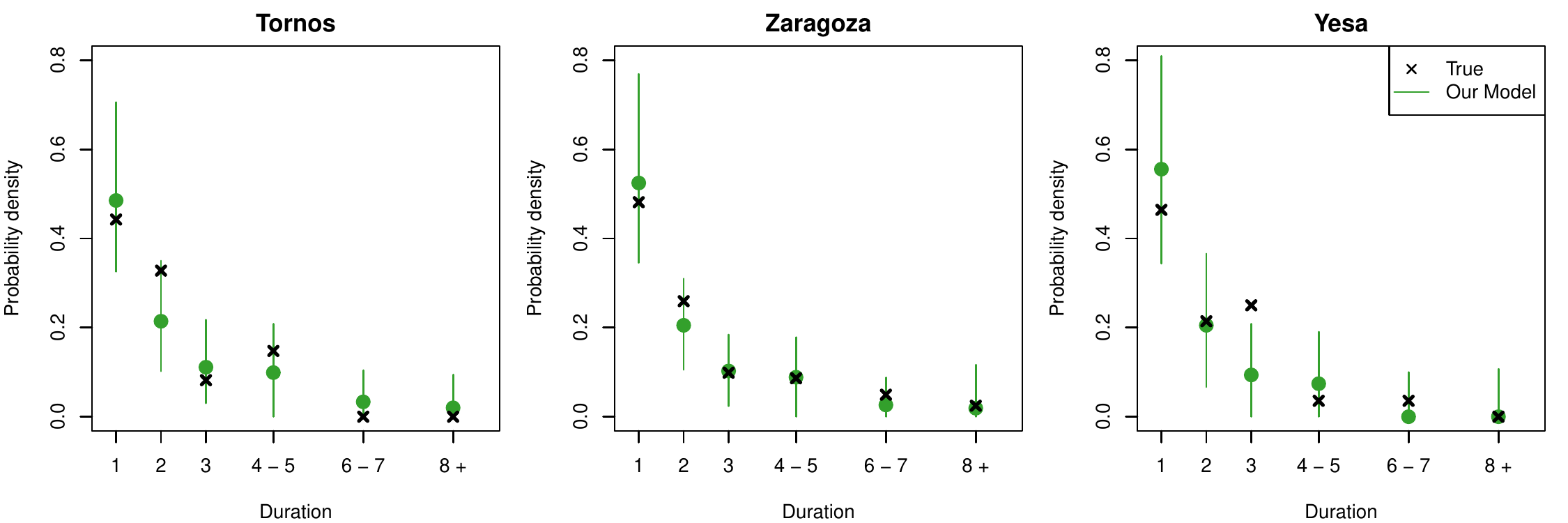}
\caption{Posterior predictive mean estimates and 90\% credible intervals of the probability density for the durations of extreme heat events across the years 1976-1985 (top) and 2006-2015 (bottom). For each of the three out-of-sample locations, true duration density is plotted for the durations 1, 2, and 3 days, 4-5 days, 6-7 days, and 8 or more days. \label{fig:DurationDistSupp}}
\end{center}
\end{figure}

Using posterior predictive samples of time series for each of the three out-of-sample locations, we compute the mean and 90\% credible interval of the probability density for events lasting 1, 2, and 3 days, 4-5 days, 6-7 days, and 8 or more days. 
That is, for each time series, we compute the proportion of EHEs during a specified time window lasting each of these durations and then compute the mean and 90\% credible intervals of these estimates over the posterior predictive time series.
These estimates are shown for each of the locations computed over the entire time window in Figure \ref{fig:DurationDist}.
The empirical probabilities are also shown for each location and duration length.
The results reveal that for each site and duration, our predictive intervals always capture the observed/true proportion.
Similar plots are included in the supplementary material for the two decades of interest, 1976-1985 and 2006-2015 (Figure \ref{fig:DurationDistSupp}) and reveal similar results. 

\begin{figure}
	\begin{center}
		\includegraphics[scale=.68]{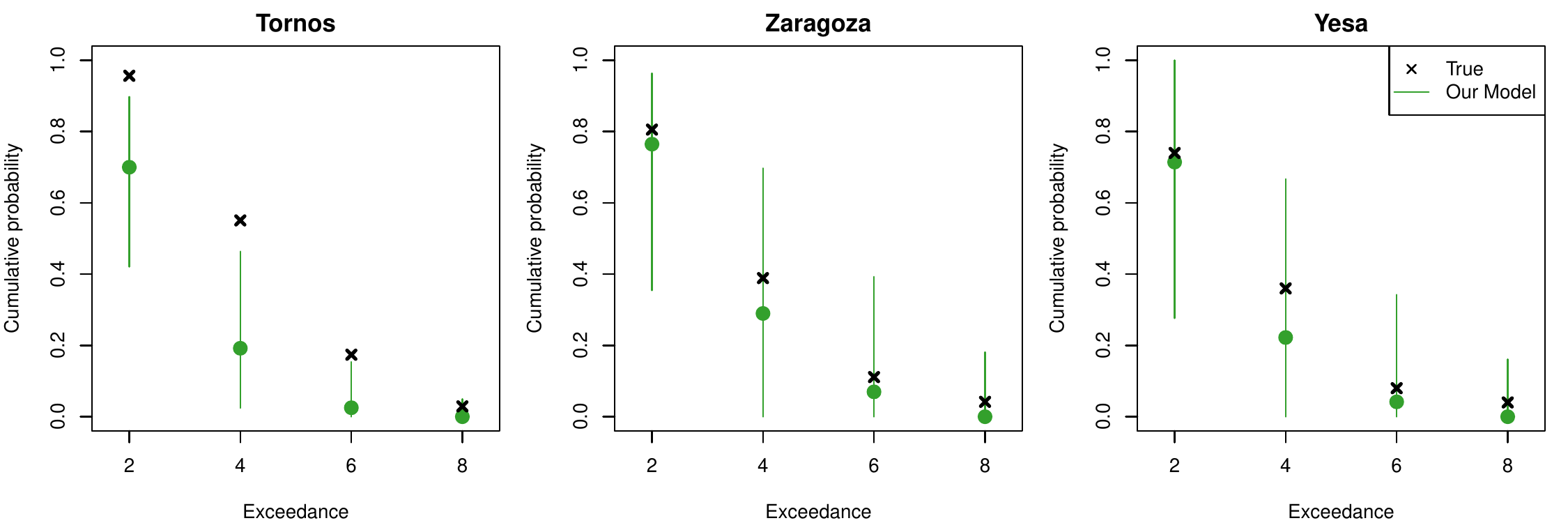}
		\includegraphics[scale=.68]{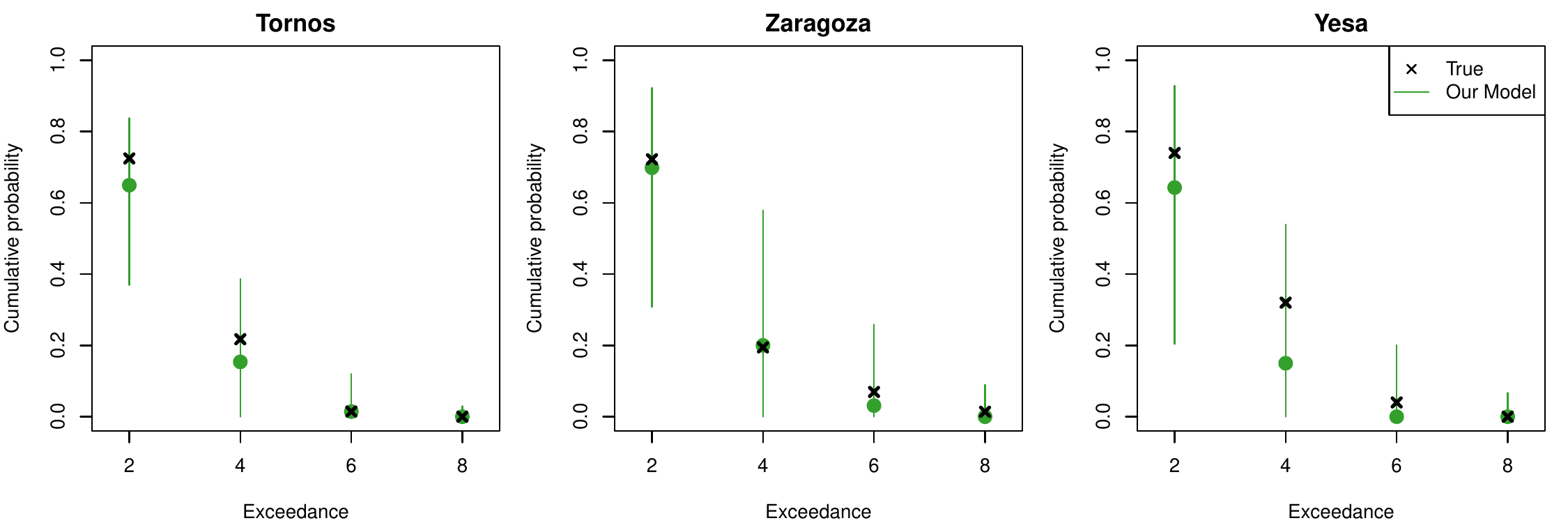}
		\caption{Posterior predictive mean estimates and 90\% credible intervals of the cumulative probability of the average (top) and maximum (bottom) exceedance being at or greater than the specified level during an EHE lasting 3 of more days.}
		 \label{fig:FeaturesMax}
	\end{center}
\end{figure}

Next, we turn to the two measures of the intensity of an EHE, the average and maximum excess (degrees over the threshold) during the event \citep{Abaurrea2007}.
Following the WMO's description of a heat wave, we focused on EHEs lasting three or more days.
For each posterior predictive time series, we computed the average (or maximum) excess for each EHE and then obtained the cumulative probability of the average (or maximum) excess being greater than or equal to a set of discrete values.
These cumulative probabilities provide an estimate of the distribution of average and maximum excess for each location.
Using all of the posterior predictive time series samples, we computed the mean and 90\% credible of these cumulative probabilities for each location.
The estimates are shown in Figure \ref{fig:FeaturesMax} for the average (top) and maximum (bottom) excess for each out of sample locations over the entire time window of the study.
Our model appears to capture these cumulative probabilities well for both average and maximum exceedance.
Similar plots are shown in Figures \ref{fig:SuppFeat1} and \ref{fig:SuppFeat2} of the supplementary material for the years 1976-1985 and 2006-2015 with similar conclusions.

\subsection{Model inference}

\subsubsection{Models for daily maximum temperature}

Both the above and below threshold processes for daily maximum temperature include global and local spatial intercepts, annual random effects, fixed effects of trigonometric terms to provide annual seasonality, spatial covariates, elevation and latitude, and an autoregressive term to model temporal autocorrelation.
The posterior distributions for the autoregressive coefficients and the coefficients of the spatial covariates are summarized by their mean and 90\% credible intervals in Table \ref{TPostmean}. The temporal dependence is similar in both models with a high autoregressive coefficient of approximately 0.7.
The elevation coefficient, which represents the gradient of temperature with respect to the elevation, is negative in both cases, but slightly smaller in magnitude for the below threshold process.
On the other hand, there is evidence of a latitudinal effect only in the below threshold process, where it reflects a temperature decrease with increasing (northern) latitudes.
The credible intervals are more precise in the below threshold process due to a larger sample size.

\begin{table}[ht]
	\caption{Posterior mean and 90\% credible intervals for the coefficients of the below and above threshold process for daily maximum temperature.}
	\label{TPostmean}
	\centering
	\begin{tabular}{lrr}
		\hline
		& \multicolumn{1}{c}{Below threshold} &\multicolumn{1}{c}{Above threshold}\\
		& \multicolumn{1}{c}{process} & \multicolumn{1}{c}{process} \\
		\hline
		Intercept 				& 18.73 (18.54, 18.91) & 22.76 (22.05, 23.40) \\
		Elevation 				& -1.65 (-1.88, -1.42) & -2.07 (-3.07, -1.01) \\
		Latitude 				& -0.98 (-1.66, -0.02) & 0.79 (-1.71, 4.54) \\
		Autoregressive coefficient & 0.73 (0.73, 0.73) & 0.71 (0.68, 0.74) \\
		\hline
	\end{tabular}
\end{table}

\begin{figure}
	\begin{center}
		\includegraphics[scale=.68]{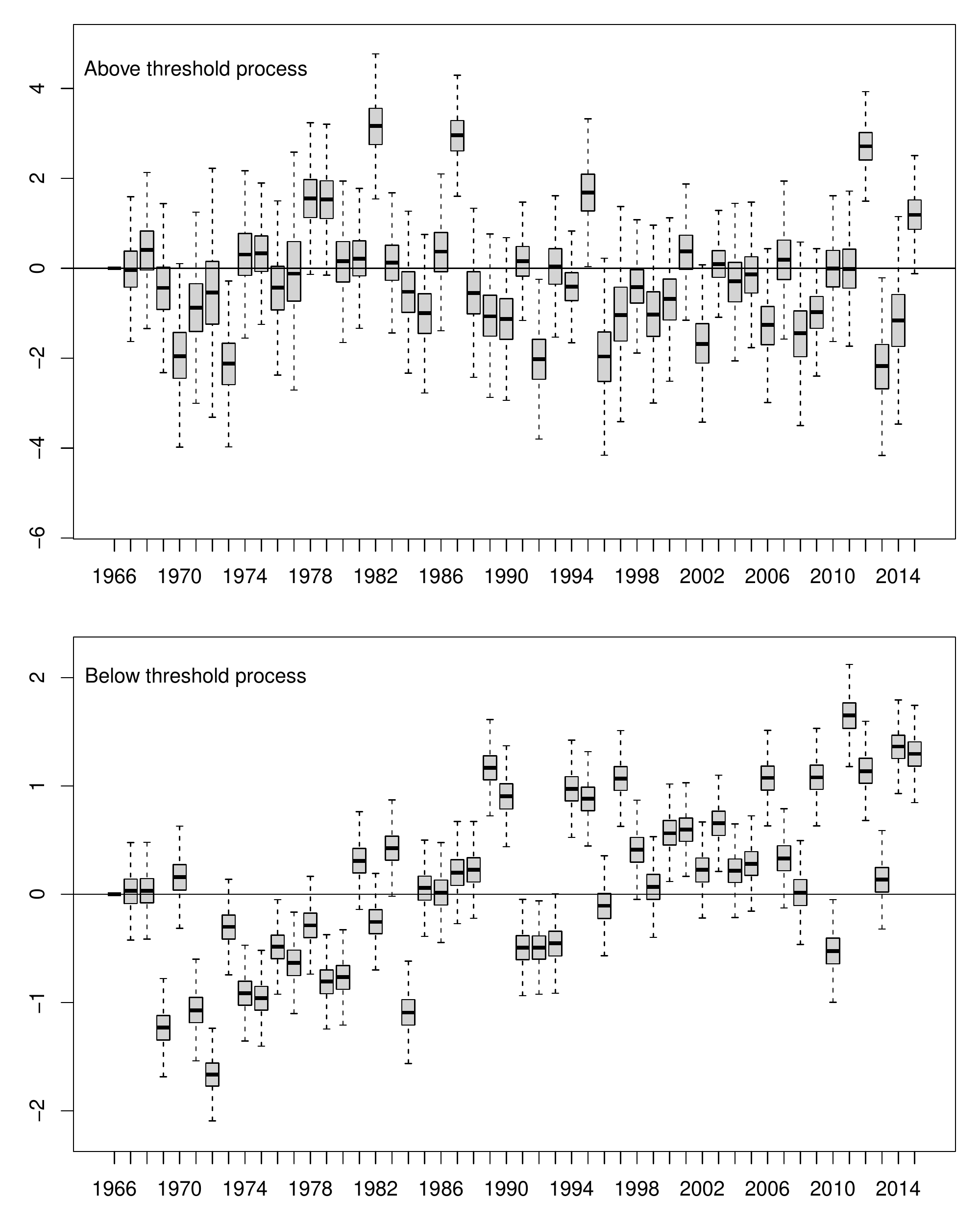}
		\caption{Boxplots of the posterior distributions of the annual random effects, $\gamma$, for the above (top) and below (bottom) threshold processes for daily maximum temperature.}
		\label{FReffects}
	\end{center}
\end{figure}

Boxplots showing the posterior distributions of the annual random effects, $\gamma$, in the above and below models are shown in Figure \ref{FReffects}.
The time evolution is quite different for the two models: while the below threshold process shows a clear increasing trend of daily maximum temperature through time, no monotonic temporal trend is detected in the above threshold process.
This signifies that the model detects an overall warming trend in daily maximum temperature between the years 1966 and 2015, yet the extreme heat temperatures are remaining relatively constant.
A similar conclusion was obtained in \cite{Abaurrea2007}, who did not find any evidence of trend in the distribution of the maximum and mean intensity of EHEs from daily temperature data obtained for the same region.
This useful inference can only be obtained from specifying different models for daily maximum temperatures above and below the threshold.

\begin{figure}
\begin{center}
\includegraphics[scale=.68]{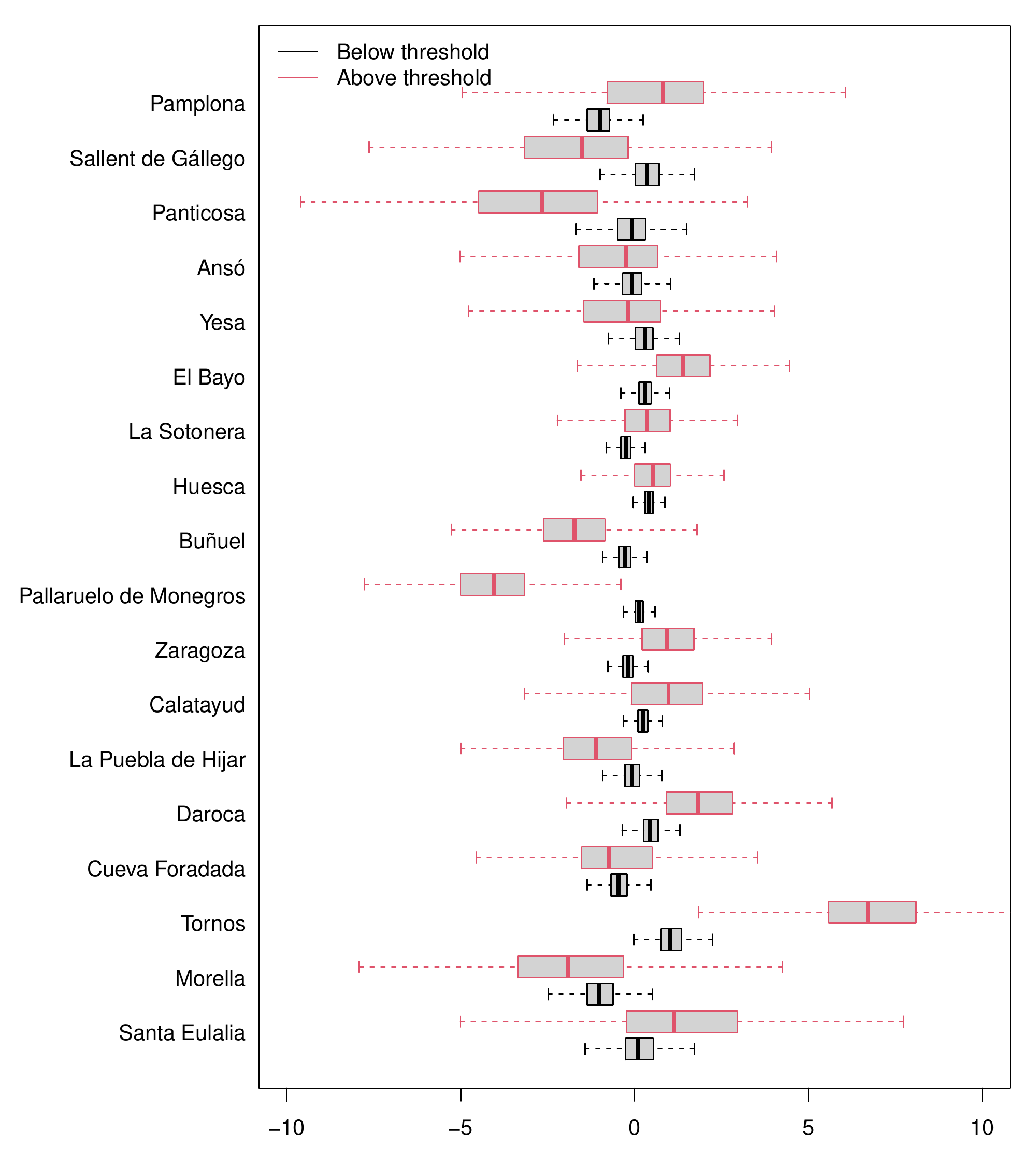}
\caption{Posterior distributions of the spatial random effects, $\beta_0(\mathbf{s})$, for the below (black) and above (red) threshold processes for daily maximum temperature. Locations are sorted latitudinally, where those at the top are in northern Arag\'{o}n and those at the bottom are in southern Arag\'{o}n. \label{Fig:Beta0}}
\end{center}
\end{figure}

The posterior distributions of the spatial random effects, $\beta_0(\mathbf{s})$, at the observed locations suggest that, in general, the spatial local adjustment to the global intercept is not very strong. Only the posterior distributions for Pallaruelo de Monegros and Tornos did not contain zero (supplementary material Figure \ref{Fig:Beta0}).
We did discover spatial differences in the standard deviations, $\sigma(\mathbf{s})$, across the region, both in the above and below threshold process, supporting the hypothesis that variation in different locations is not the same. Additionally, the mean of the spatial processes of the above and below threshold standard deviations had posterior means of 1.56 and 2.75, respectively. This is in accord with our exploratory data analysis in Section \ref{Sec:Data}.
Boxplots of these posterior distributions are given in  the supplementary material (Figure \ref{Fig:Sigma}).  

\subsubsection{The above and below threshold state models}

The models for the above and for the below threshold states include a global and local spatial intercept, fixed effects for the difference between the previous temperature and the threshold, and trigonometric terms.
The posterior distributions of the spatial random effects, $\phi_0(\mathbf{s})$, show large spatial variation in the model for the above and below threshold states.
Boxplots of these posterior distributions are shown in Figure \ref{Fig:phi0} of the supplementary material.

\begin{figure}
	\begin{center}
		\includegraphics[scale=.68]{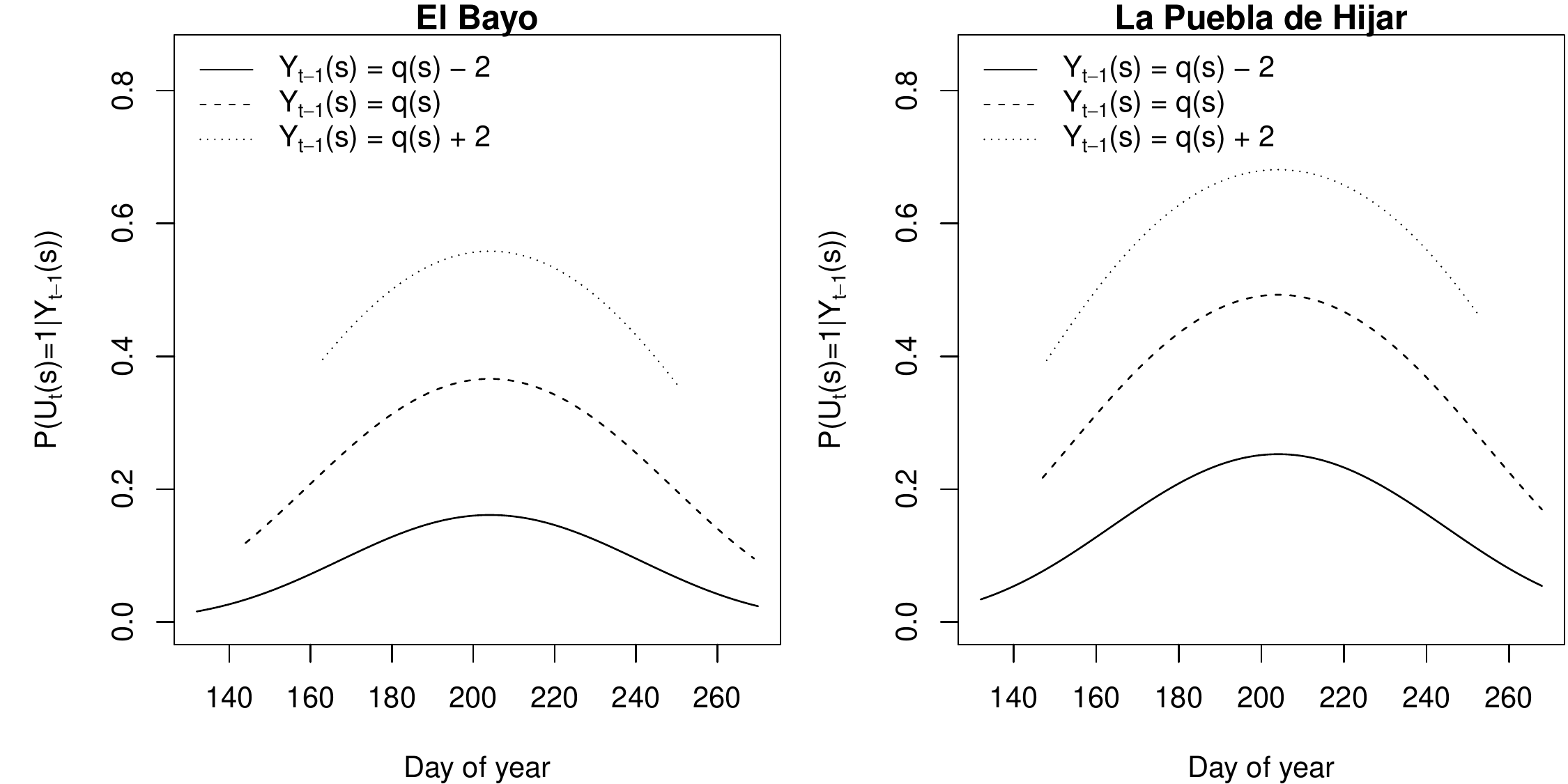}
		\caption{Posterior mean estimates of $P(U_t(\mathbf{s}) =1|Y_{t-1}(\mathbf{s}))$ across day of the year for three values of $Y_{t-1}(\mathbf{s})$. Curves are shown between the minimum and maximum day of the year in which the temperature was observed at or above the given value of $Y_{t-1}(\mathbf{s})$. }
		\label{Fig:Lambda}
	\end{center}
\end{figure}

The posterior mean estimates of the conditional probabilities $P(U_t(\mathbf{s}) =1|Y_{t-1}(\mathbf{s}))$ across day of the year for three values of $Y_{t-1}(\mathbf{s})$ corresponding to a previous day with temperature below, equal to, and above the threshold, are shown in Figure \ref{Fig:Lambda}. These conditional probabilities are shown for El Bayo (left) and La Puebla de Hijar (right), as these locations have the most extreme random effects for the local intercept $\phi_0(\mathbf{s})$.
Since the model is specified using $Y_{t-1}(\mathbf{s}) -q(\mathbf{s})$ as opposed to raw temperature values, we do not expect to see major differences between these distributions, even though the climate of these two locations are different.
However, we do see larger probabilities for La Puebla de Hijar than El Bayo for all three values of $Y_{t-1}(\mathbf{s})$ indicating a greater likelihood for EHE events to occur or persist.
The mean estimates of the probabilities show a clear seasonal behavior, modeled by the trigonometric terms, with the maximum value occurring on day 203 (July 22, July 21 in leap years).  
The effect of the previous daily temperature (relative to the threshold) on the probability of being over the threshold is also very strong. For example, the probability in the middle of summer when $Y_{t-1}(\mathbf{s}) = q(\mathbf{s}) + 2^{\text{o}}$C  is three times the probability when $Y_{t-1}(\mathbf{s}) = q(\mathbf{s}) - 2^{\text{o}}$C.
This difference is evidence of the increased probability of persistence compared to the probability of the first exceedance day of an EHE.

\section{Discussion and future work}
\label{Sec:Disc}

We have proposed a novel threshold-driven spatial autoregressive two-state model for learning about features of extreme heat events over a $60+$ year period for the region of Arag\'{o}n in Spain. These features include incidence, duration, maximum of the daily maximum temperatures over the event, and average exceedance above the threshold over the event. The two-state model captures tail dependence for exceedances using truncated t-distributions with normal distributions for temperatures below thresholds.  Using leave one out validation, we demonstrate that the two-state model predicts these features well.

Future work includes the possibility of comparison of results for the region of Arag\'{o}n with other regions in Spain since national temperature databases are available. In this regard, we could consider a national assessment of EHE characteristics. This extension would likely require the inclusion of spatially-varying coefficients, anticipating that, e.g., mountainous response to predictors would be different from coastal response to the predictors.  
We might also consider seasonal variation in uncertainty, motivating the need for time-varying variances beyond our current above and below spatially varying specifications.  Finally, we plan to use our approach to forecasting future EHE behavior. Using regional climate model scenarios over 50 year windows, these forecasts could provide useful insight into the evolution of EHE behavior over the future time period.

\bibliographystyle{apalike}
\bibliography{EHEReferencesSpatial}
\paragraph{Acknowledgements}
Authors Abaurrea, As\'{i}n, Beamonte, and Cebri\'{a}n are members of the research group Modelos Estoc\'{a}sticos, supported by the DGA (Government of Arag\'{o}n), the European Social Fund and the project MTM2017-83812-P. The authors acknowledge AEMET (Spanish Agency of Meteorology) for the data. Lastly, we thank the reviewers and associate editor for their thoughtful comments that greatly improved our manuscript.

\nolinenumbers
\newpage
\section*{Supplementary Material}
\beginsupplement
\setcounter{page}{1}


\begin{figure}[h!]
\begin{center}
\includegraphics[scale=.53]{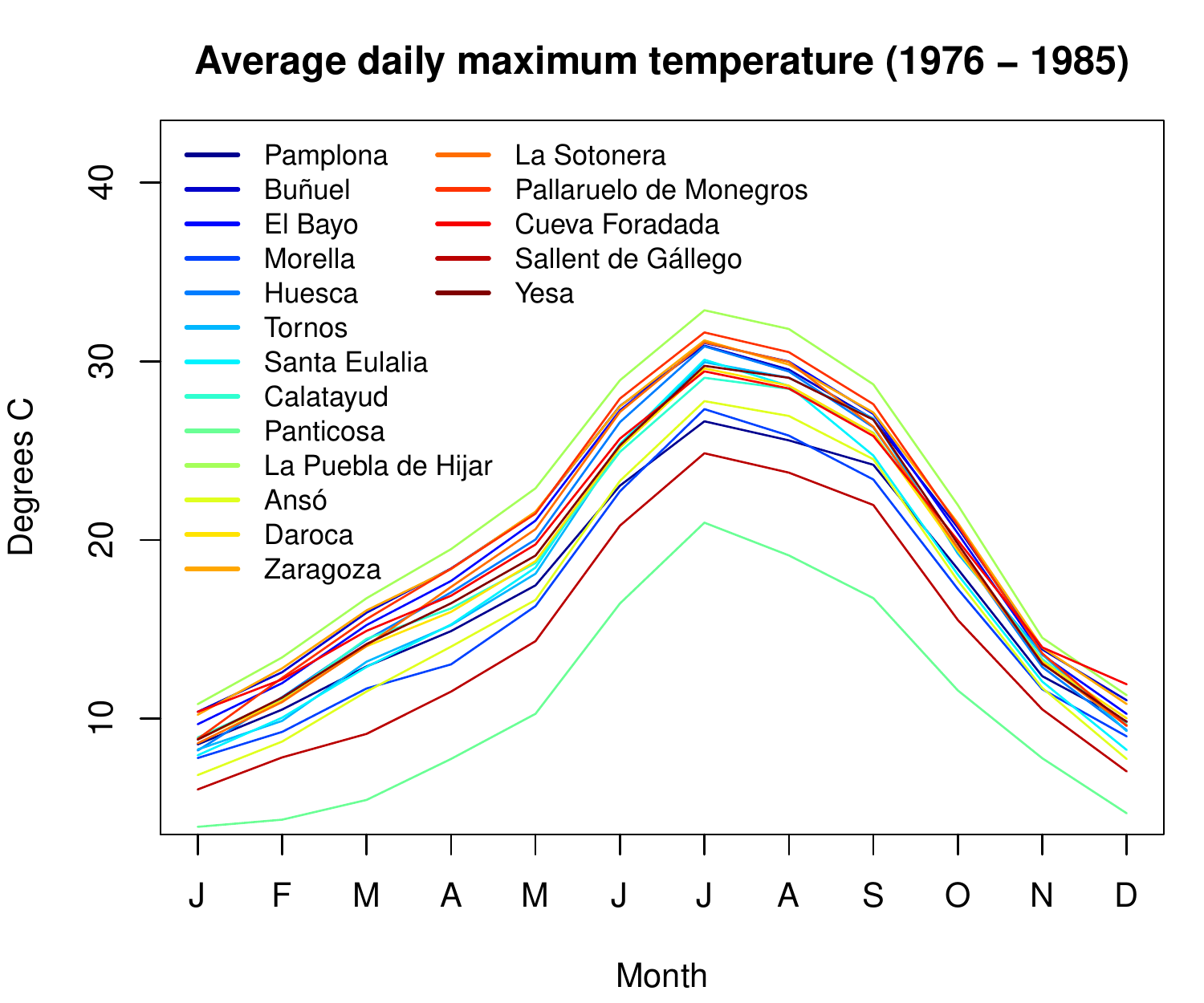}
\includegraphics[scale=.53]{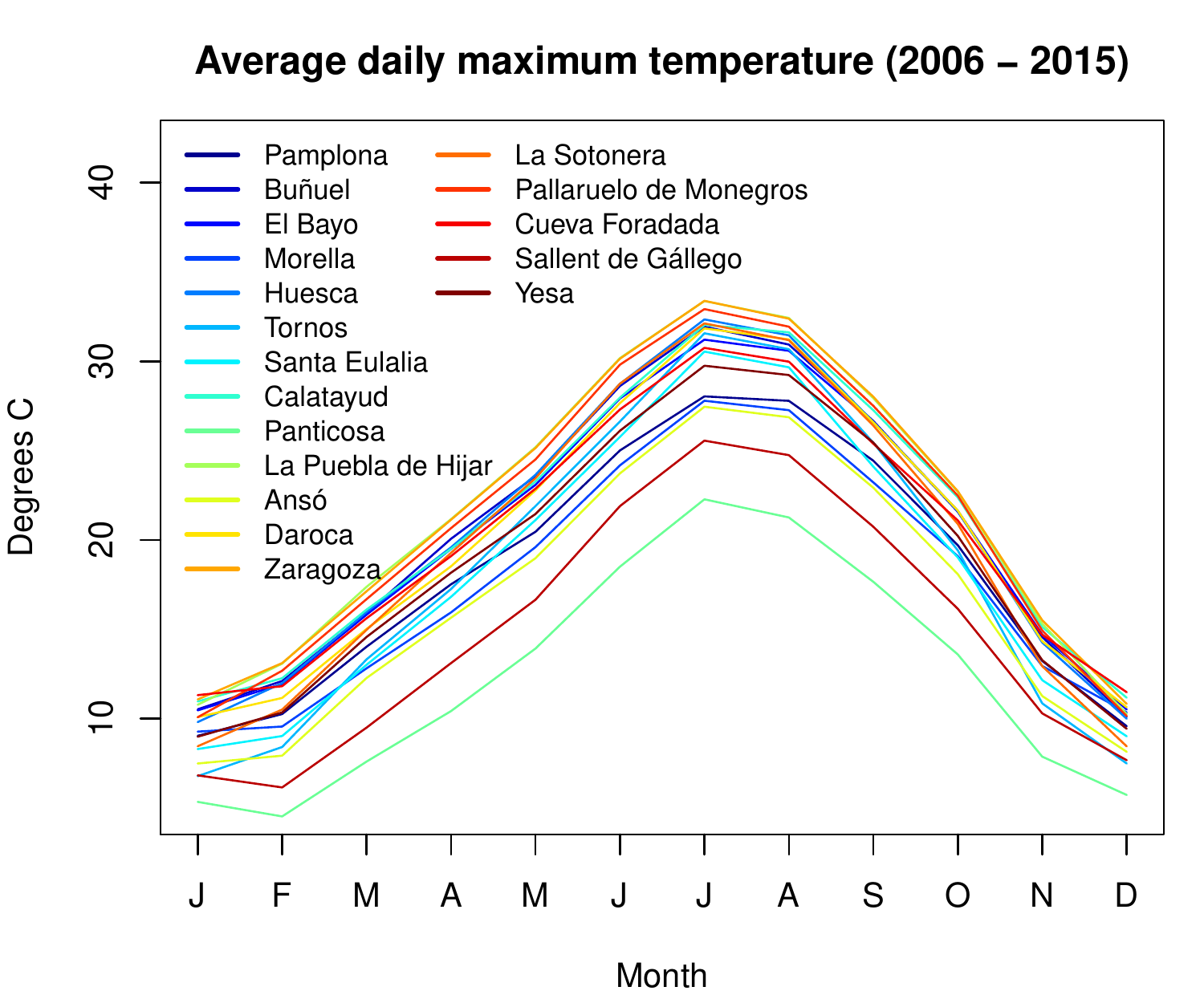}\\
\includegraphics[scale=.53]{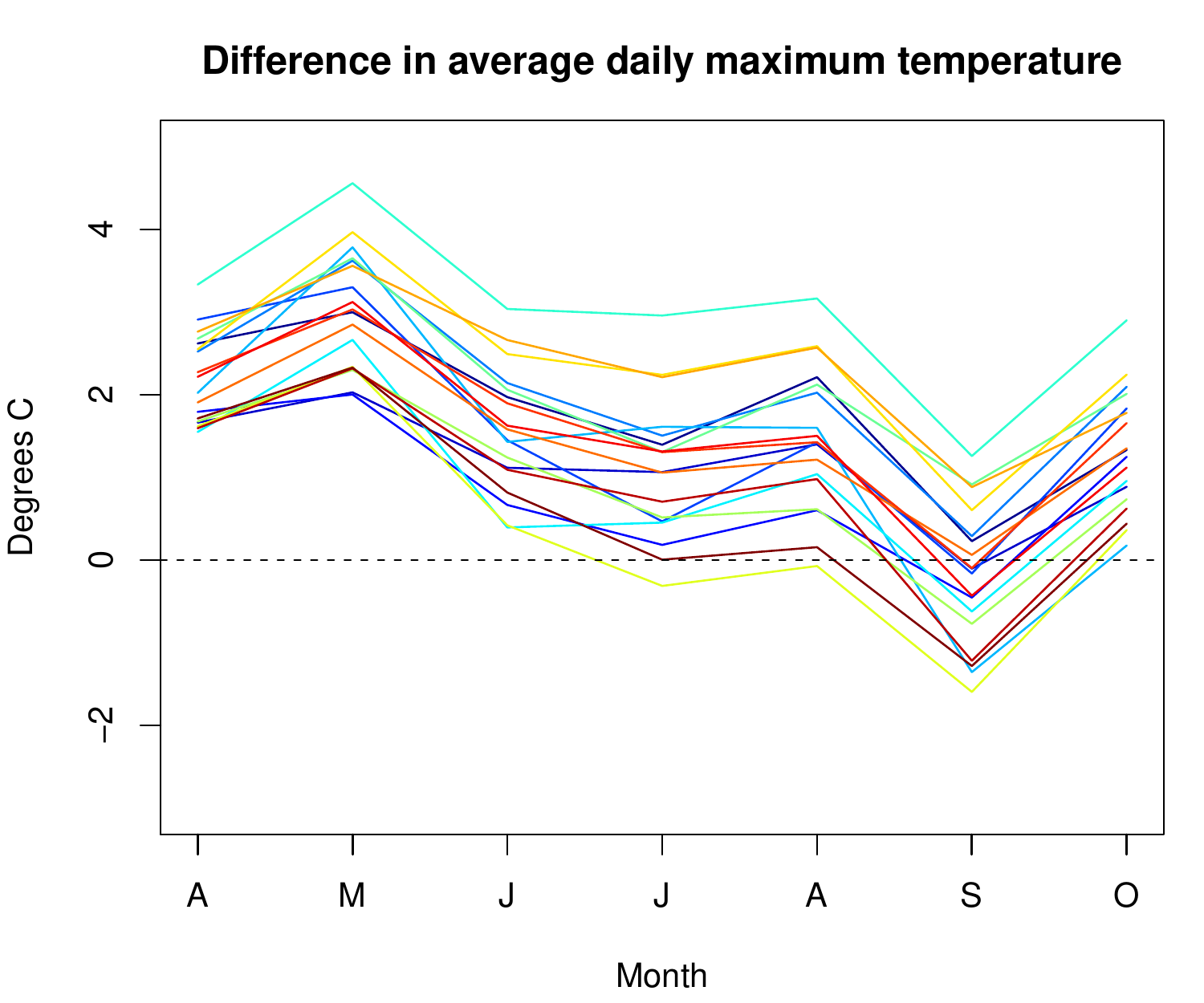}
\caption{Monthly average daily maximum temperature for each location during for the years 1976-1985 (top, left) and 2006-2015 (top, right), as well as the difference in these monthly averages between the two time periods for the months April through October (bottom). \label{fig:TS}}
\end{center}
\end{figure}

\begin{figure}
	\begin{center}
		\includegraphics[scale=.68]{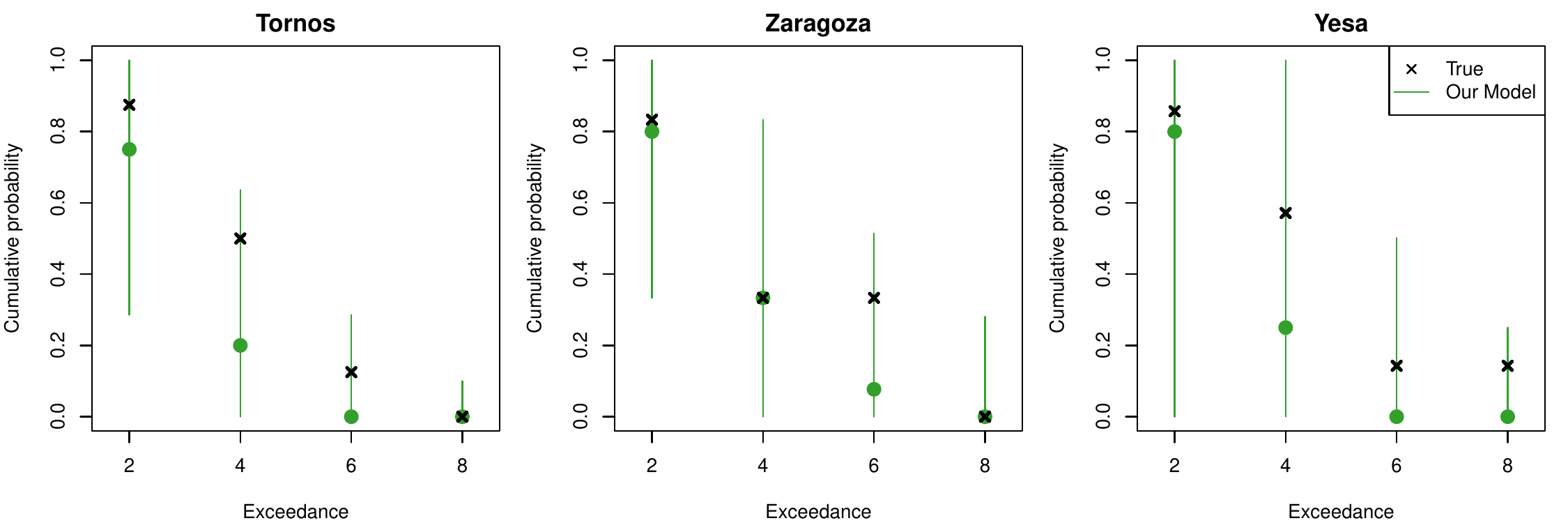}
		\includegraphics[scale=.68]{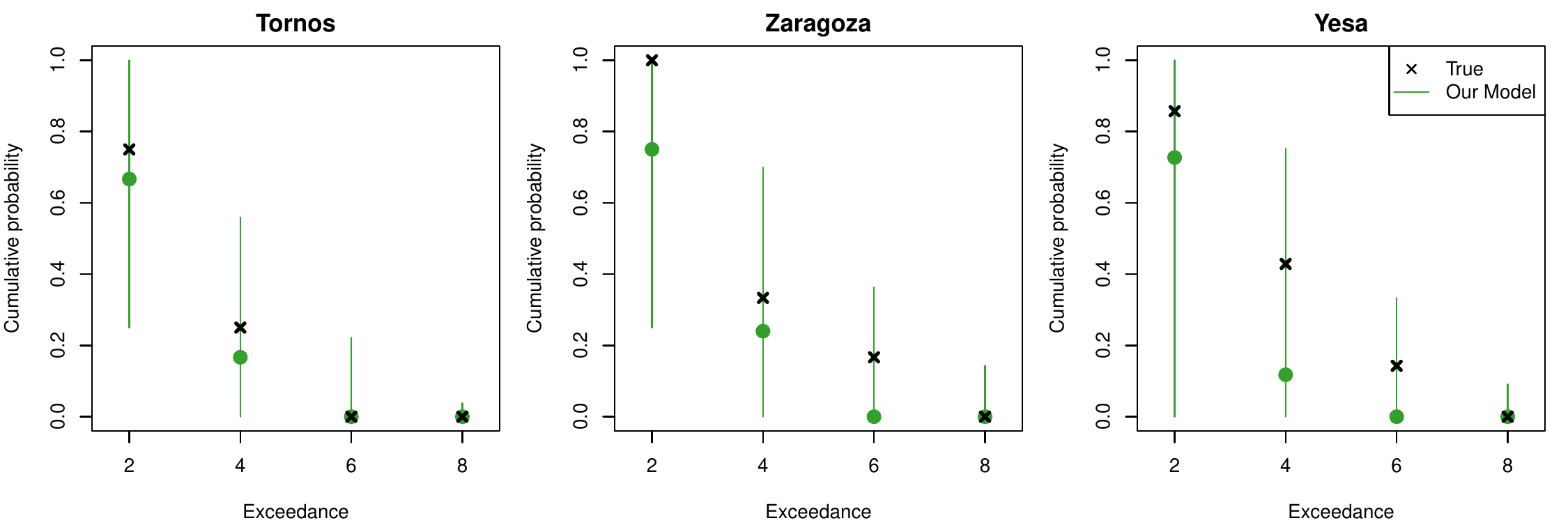}
		\caption{Posterior predictive mean estimates and 90\% credible intervals of the cumulative probability of the average (top) and maximum (bottom) exceedance being at or greater than the specified level during an EHE lasting 3 of more days during the years 1976-1985. x denotes the observed value.}
		 \label{fig:SuppFeat1}
	\end{center}
\end{figure}

\begin{figure}
	\begin{center}
		\includegraphics[scale=.68]{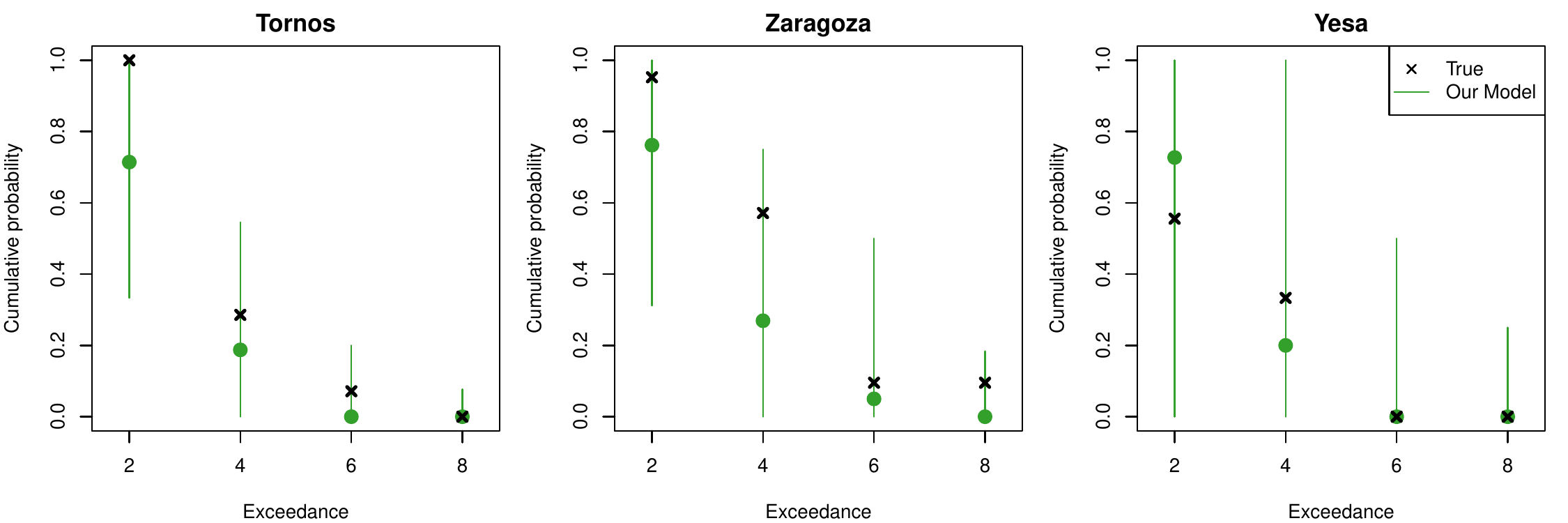}
		\includegraphics[scale=.68]{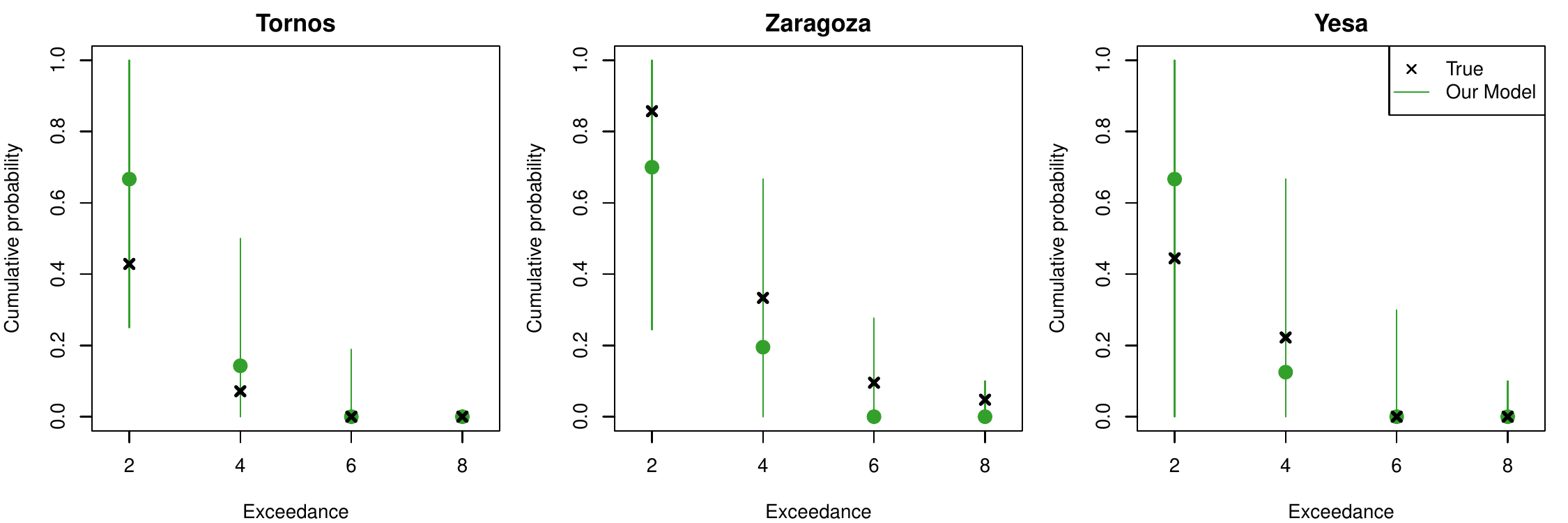}
		\caption{Posterior predictive mean estimates and 90\% credible intervals of the cumulative probability of the average (top) and maximum (bottom) exceedance being at or greater than the specified level during an EHE lasting 3 of more days during the years 2006-2015. x denotes the observed value.}
		 \label{fig:SuppFeat2}
	\end{center}
\end{figure}

\begin{figure}
\begin{center}
\includegraphics[scale=.68]{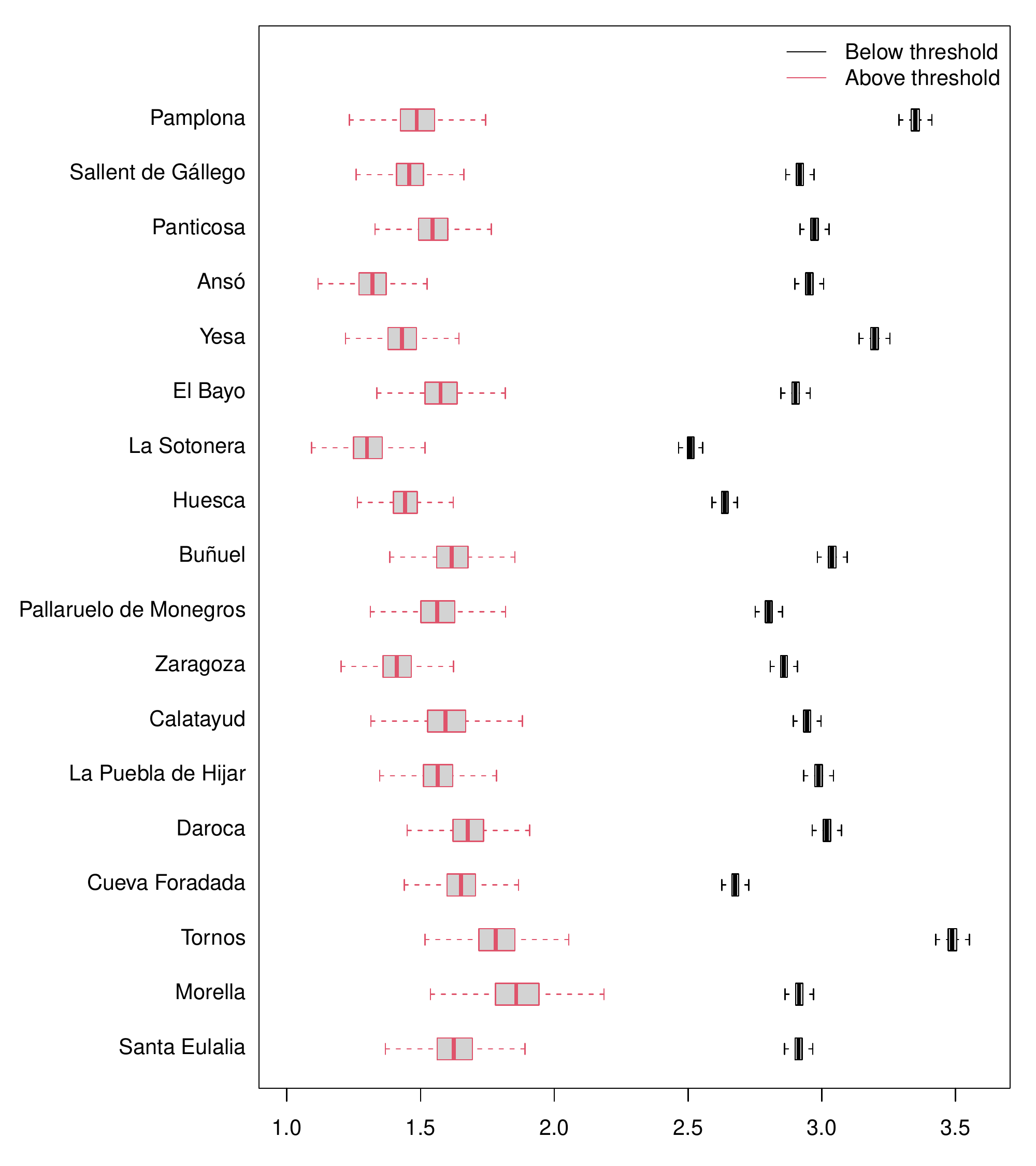}
\caption{Posterior distributions of the spatial standard deviations, $\sigma(\mathbf{s})$, for the below (black) and above (red) threshold processes for daily maximum temperature. Locations are sorted latitudinally, from north to south. \label{Fig:Sigma}}
\end{center}
\end{figure}

\begin{figure}
\begin{center}
\includegraphics[scale=.68]{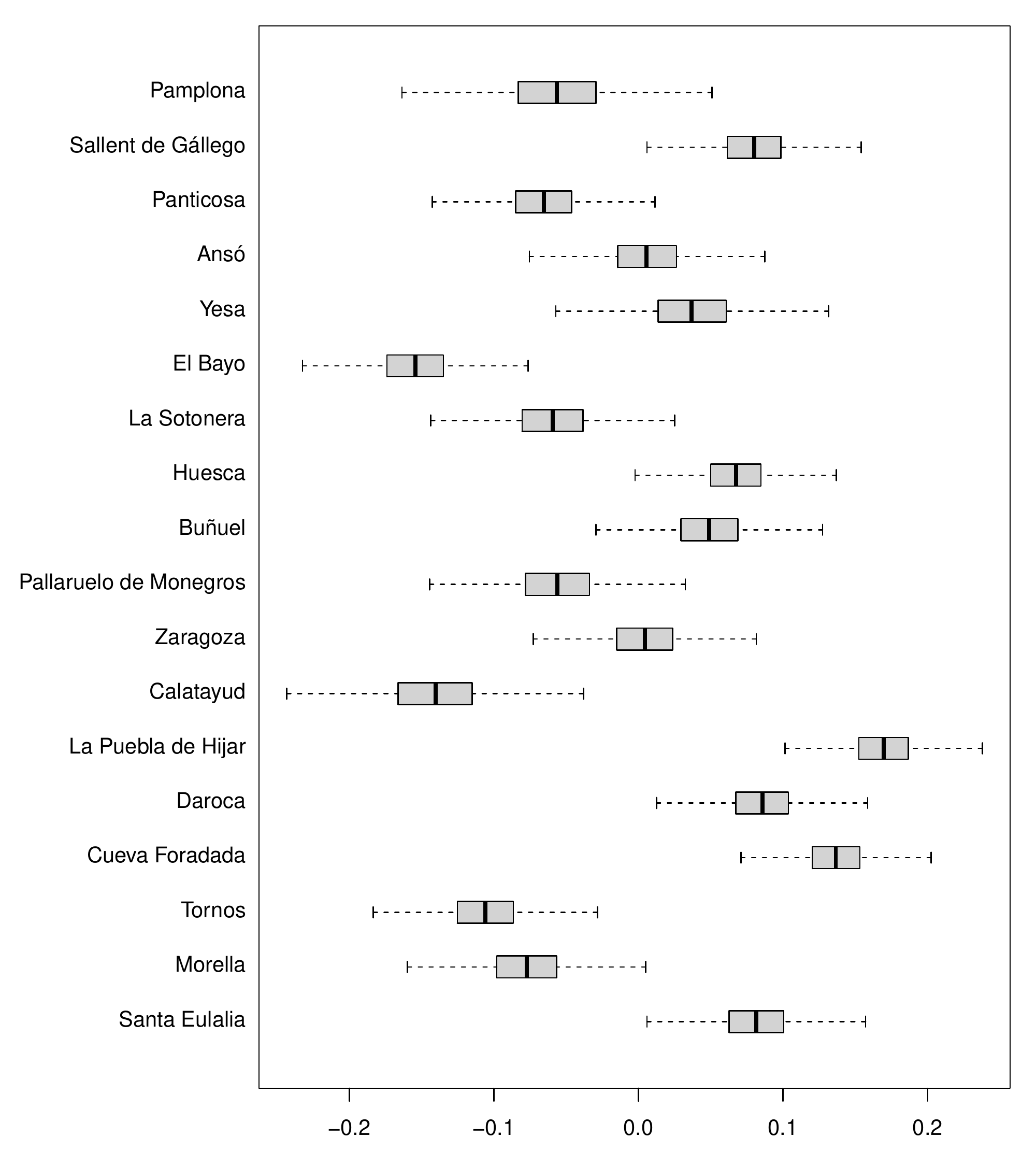}
\caption{Boxplots of the posterior distributions of the spatial random effects, $\phi_{0}(\bs)$ of the $U$ process. Locations are sorted latitudinally, from north to south. \label{Fig:phi0}}
\end{center}
\end{figure}


\end{document}